\documentclass[prd,tightenlines,showpacs,nofootinbib,eqsecnum,amsfonts,amssymb,amsmath,twocolumn,widetext]{revtex4}

\usepackage{graphicx}
\usepackage{amsmath}
\usepackage{amsfonts} 
\usepackage{amssymb}  
\usepackage{mathrsfs}
\usepackage{epsfig,bbm}
\usepackage{bm}
\usepackage{color}
\usepackage{dsfont}
\usepackage{tensor}
\usepackage{microtype}

\usepackage{wasysym}
\usepackage{amsthm}
\usepackage{hyperref}

\begin{document}


\newcommand{\average}[1]{\langle{#1}\rangle_{{\cal D}}}
\newcommand{\dd}{{\rm d}}
\newcommand{\etal}{{\it et al.}}
\newcommand{\e}[1]{_{\text{#1}}}
\newcommand{\h}[1]{^{\text{#1}}}
\newcommand{\ph}{\varphi}
\newcommand{\eps}{\varepsilon}
\newcommand{\hreftt}[2]{\href{#1}{\texttt{#2}}}


\title{Interpretation of the Hubble diagram in a nonhomogeneous universe}

\author{Pierre Fleury$^{1,2}$}
\email{fleury@iap.fr}       
             
\author{H\'el\`ene Dupuy$^{1,2,3}$}
\email{helene.dupuy@cea.fr}         

\author{Jean-Philippe Uzan$^{1,2}$}
\email{uzan@iap.fr}
 
\affiliation{
$^1$ Institut d'Astrophysique de Paris, UMR-7095 du CNRS, Universit\'e Pierre et Marie Curie, 98 bis bd Arago, 75014 Paris, France.\\
$^2$ Sorbonne Universit\'es, Institut Lagrange de Paris, 98 bis bd Arago, 75014 Paris, France.\\
$^3$ Institut de Physique Th\'eorique,
	 CEA, IPhT, URA 2306 CNRS, F-91191 Gif-sur-Yvette, France.
}

\begin{abstract}
In the standard cosmological framework, the Hubble diagram is interpreted by assuming that the light emitted by standard candles propagates in a spatially homogeneous and isotropic spacetime. However, the light from ``point sources''---such as supernovae---probes the Universe on scales where the homogeneity principle is no longer valid. Inhomogeneities are expected to induce a bias and a dispersion of the Hubble diagram. This is investigated by considering a Swiss-cheese cosmological model, which (1) is an exact solution of the Einstein field equations, (2) is strongly inhomogeneous on small scales, but (3) has the same expansion history as a strictly homogeneous and isotropic universe. By simulating Hubble diagrams in such models, we quantify the influence of inhomogeneities on the measurement of the cosmological parameters. Though significant in general, the effects reduce drastically for a universe dominated by the cosmological constant.
\end{abstract}

\date{\today}

\maketitle

\section{Introduction}
\label{sec:introduction}

The standard physical model of cosmology relies on a solution of general relativity describing a spatially homogeneous and isotropic spacetime, known as the Friedmann-Lema\^{\i}tre (FL) solution (see e.g. Ref.~\cite{pubook}). It is assumed to describe the geometry of our Universe smoothed on large scales. Besides, the use of the perturbation theory allows one to understand the properties of the large scale structure, as well as its growth from initial conditions set by inflation and constrained by the observation of the cosmic microwave background.

While this simple solution of the Einstein field equations, together with the perturbation theory, provides a description of the Universe in agreement with all existing data, it raises many questions on the reason why it actually gives such a good description. In particular, it involves a smoothing scale which is not included in the model itself~\cite{r.scale}. This opened a lively debate on the fitting problem~\cite{r.fitting} (i.e. what is the best-fit FL model to the lumpy Universe?) and on backreaction (i.e. the fact that local inhomogeneities may affect the cosmological dynamics). The amplitude of backreaction is still actively debated~\cite{buchertetal,nobackreac,backsig}, see Ref.~\cite{review-backreac} for a critical review. 

Regardless of backreaction, the cosmological model assumes that the distribution of matter is continuous (i.e. it assumes that the fluid approximation holds on the scales of interest) both at the background and perturbation levels. Indeed numerical simulations fill part of this gap by dealing with $N$-body gravitational systems in an expanding space. The fact that matter is not continuously distributed can however imprint some observations, in particular regarding the propagation of light with narrow beams, as discussed in detail in Ref.~\cite{cemuu}. It was argued that such beams, as e.g. for supernova observations, probe the spacetime structure on scales much smaller than those accessible in numerical simulations. The importance of quantifying the effects of inhomogeneities on light propagation was first pointed out by Zel'dovich~\cite{zel}. Arguing that photons should mostly propagate in vacuum, he designed an ``empty beam'' approximation, generalized later by Dyer and Roeder as the ``partially filled beam'' approach~\cite{DR72}. More generally, the early work of Ref.~\cite{zel} stimulated many studies on this issue \cite{dash,Bertotti,gunn,refs,weinberg,DR81,fangwu,Rose:2001qi,
Kibble:2004tm,Kostov:2009uc,linder88,tomita,Mortsell:2001es,
Bolejko:2010nh,Takahashi:2011qd}.

The propagation of light in an inhomogeneous universe gives rise to both distortion and magnification induced by gravitational lensing. While most images are demagnified, because most lines of sight probe underdense regions, some are amplified because of strong lensing. Lensing can thus discriminate between a diffuse, smooth component, and the one of a gas of macroscopic, massive objects (this property has been used to probe the nature of dark matter~\cite{sh99,Metcalf1999,Metcalf2006}). Therefore, it is expected that lensing shall induce a dispersion of the luminosities of the sources, and thus an extra scatter in the Hubble diagram~\cite{scatterHubble}. Indeed, such an effect does also appear at the perturbation level---i.e. with light propagating in a perturbed FL spacetime---and it was investigated in Refs.~\cite{Valageas2000,Bonvin:2005ps,Meures:2011gp,Marozzi2012,DiDio2012,Marozzi2013}. The dispersion due to the large-scale structure becomes comparable to the intrinsic dispersion for redshifts $z > 1$~\cite{hl} but this dispersion can actually be corrected~\cite{Cooray:2005yp,Cooray:2005yr,Sarkar:2007sp,Cooray:2008qn,Vallinotto:2010qm,menard03}. Nevertheless, a considerable fraction of the lensing dispersion arises from sub-arc minute scales, which are not probed by shear maps smoothed on arc minute scales~\cite{Dalal2002}. The typical angular size of the light beam associated with a supernova (SN) is typically of order $10^{-7}$~arc sec (e.g. for a source of physical size $\sim1$ AU at redshift~$z\sim1$), while the typical observational aperture is of order  1~arc sec. This is smaller than the mean distance between any massive objects.

One can estimate~\cite{Metcalf1999} that a gas composed of particles of mass $M$ can be considered diffuse on the scale of the beam of an observed source of size~$\lambda_s$ if $M<2\times 10^{-23} M_\odot h^2 \left({\lambda_{\rm s}}/{1\,\text{AU}}\right)^3$. In the extreme case for which matter is composed only of macroscopic pointlike objects, then most high-redshift SNeIa would appear fainter than in a universe with the same density distributed smoothly, with some very rare events of magnified SNeIa~\cite{Metcalf1999,rauch,HW97}. This makes explicit the connection between the Hubble diagram and the fluid approximation which underpins its standard interpretation.

The fluid approximation was first tackled in a very innovative work of Lindquist and Wheeler~\cite{wheeler}, using a Schwarzschild cell method modeling an expanding universe with spherical spatial sections. For simplicity, they used a regular lattice which restricts the possibilities to the most homogeneous topologies of the 3-sphere~\cite{topos3}. It has recently been revisited in Refs.~\cite{archipele} and in Refs.~\cite{larena} for Euclidean spatial sections. They both constructed the associated Hubble diagrams, but their spacetimes are only approximate solutions of the Einstein field equations. An attempt to describe filaments and voids was also proposed in Ref.~\cite{KM:2009}.

These approaches are conceptually different from the solution we adopt in the present article. We consider an exact solution of the Einstein field equations with strong density fluctuations, but which keeps a well-defined FL averaged behavior. Such conditions are satisfied by the Swiss-cheese model~\cite{EinStr45}: one starts with a spatially homogeneous and isotropic FL geometry, and then cuts out spherical vacuoles in which individual masses are embedded. Thus, the masses are contained in vacua within a spatially homogeneous fluid-filled cosmos (see bottom panel of Fig.~\ref{fig:FL_VS_SC}). By construction, this exact solution is free from any backreaction: its cosmic dynamics is identical to the one of the underlying FL spacetime.

\begin{figure}[h!]
\centering
\includegraphics[width=\columnwidth]{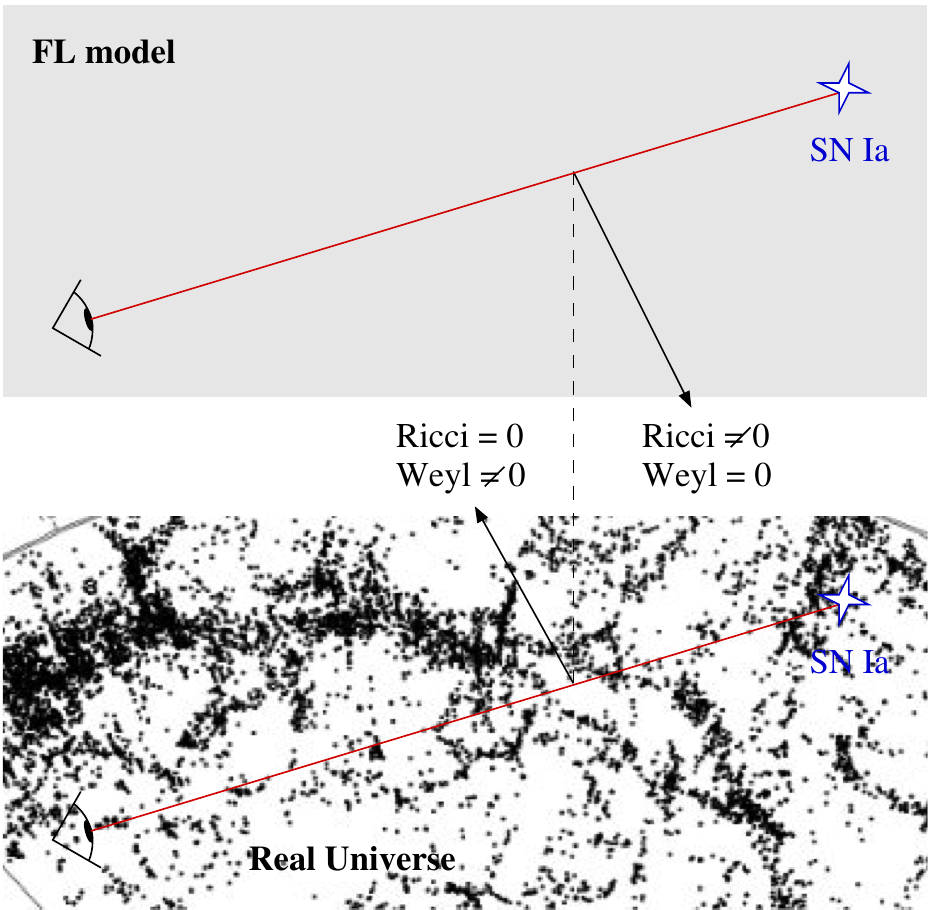}
\caption{The standard interpretation of SNe data assumes that light propagates in purely homogeneous and isotropic space~(top). However, thin light beams are expected to probe the inhomogeneous nature of the actual Universe (bottom) down to a scale where the continuous limit is no longer valid.
}
\label{fig:FL_VS_reality}
\end{figure}

\begin{figure}[h!]
\centering
\includegraphics[width=\columnwidth]{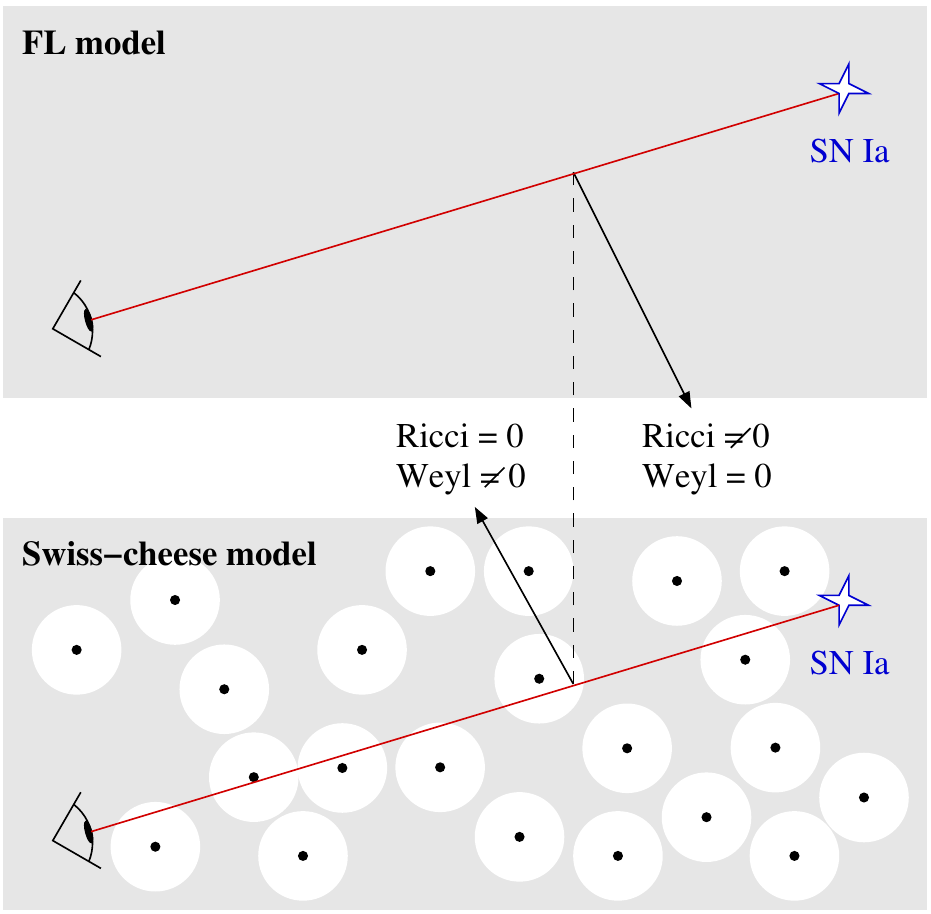}
\caption{Swiss-cheese models (bottom) allow us to model inhomogeneities beyond the continuous limit, while keeping the same dynamics and average properties as the FL model (top).}
\label{fig:FL_VS_SC}
\end{figure}

From the kinematical point of view, Swiss-cheese models allow us to go further than perturbation theory, because not only the density of matter exhibits finite fluctuations, but also the metric itself. Hence, light propagation is expected to be very different in a Swiss-cheese universe compared to its underlying FL model. Moreover, the inhomogeneities of a Swiss cheese are introduced in a way that addresses the so-called ``Ricci-Weyl problem.'' Indeed, the standard FL geometry is characterized by a vanishing Weyl tensor and a nonzero Ricci tensor, while in reality light mostly travels in vacuum, where conversely the Ricci tensor vanishes---apart from the contribution of $\Lambda$, which does not focus light---and the Weyl tensor is nonzero (see Fig.~\ref{fig:FL_VS_reality}). A Swiss-cheese model is closer to the latter situation, because the Ricci tensor is zero inside the holes (see Fig.~\ref{fig:FL_VS_SC}). It is therefore hoped to capture the relevant optical properties of the Universe.

In fact, neither a Friedmann-Lema\^{\i}tre model nor a Swiss-cheese model can be considered a realistic description of the Universe. They share the property of being exact solutions of the Einstein equations which satisfy the Copernican principle, either strictly or statistically. Swiss-cheese models can be characterized by an extra-cosmological parameter describing the smoothness of their distribution of matter. Thus, a FL spacetime is nothing but a perfectly smooth Swiss cheese. It is legitimate to investigate to which extent observations can constrain the smoothness cosmological parameter, and therefore to quantify how close to a FL model the actual Universe is.

The propagation of light in a Swiss-cheese universe was first investigated by Kantowski~\cite{Kan69}, and later by Dyer and Roeder \cite{DR74}. Both concluded that the effect, on the Hubble diagram, of introducing ``clumps'' of matter was to lower the apparent deceleration parameter. The issue was revived within the backreaction and averaging debates, and the Swiss-cheese models have been extended to allow for more generic distributions of matter inside the holes---instead of just concentrating it at the center---where spacetime geometry is described by the Lema\^itre-Tolman-Bondi (LTB) solution. The optical properties of such models have been extensively studied (see Refs. \cite{Brouzakis:2006dj,Biswas:2007gi,Marra:2007pm,
Vanderveld:2008vi,Brouzakis:2008,Vanderveld:2011,Clifton:2009nv,
Valkenburg:2009iw,Szybka:2010ky}) to finally conclude that the average luminosity-redshift relation remains unchanged with respect to the purely homogeneous case, contrary to the early results of Refs.~\cite{Kan69,DR74}.

In general, the relevance of ``LTB holes'' in Swiss-cheese models is justified by the fact that they allow one to reproduce the actual large-scale structure of the Universe (with voids and walls). However, though inhomogeneous, the distribution of matter in this class of models remains continuous at all scales. On the contrary, the old-fashioned approach with ``clumps'' of matter inside the holes \emph{breaks the continuous limit}. Hence, it seems more relevant for describing the small-scale structure probed by thin light~beams.

In this article, we revisit and update the studies of Refs.~\cite{Kan69,DR74} within the paradigm of modern cosmology. For that purpose, we first provide a comprehensive study of light propagation in the same class of Swiss-cheese models, including the cosmological constant. By generating mock Hubble diagrams, we then show that the inhomogeneities induce a significant bias in the apparent luminosity-redshift relation, which affects the determination of the cosmological parameters. As we shall see, the effect increases with the fraction of clustered matter but decreases with~$\Lambda$. For a universe apparently dominated by dark energy, the difference turns out to be small.  

The article is organized as follows. Section~\ref{sec:Swiss-cheese} describes the construction and mathematical properties of the Swiss-cheese model. In Sec.~\ref{sec:geometric_optics}, we summarize the laws of light propagation, and introduce a new tool to deal with a patchwork of spacetimes, based on matrix multiplications. In Sec.~\ref{sec:resolution}, we apply the laws introduced in Sec.~\ref{sec:geometric_optics} to Swiss-cheese models and solve the associated equations. The results enable us to investigate the effect of one hole (Sec.~\ref{sec:results_one_hole}) and of many holes (Sec.~\ref{sec:results_several_holes}) on cosmological observables, namely the redshift and the luminosity distance. Finally, the consequences on the determination of the cosmological parameters are presented in Sec.~\ref{sec:cosmological_consequences}.

\section{Description of the Swiss-cheese cosmological model}
\label{sec:Swiss-cheese}

The construction of Swiss-cheese models is based on the Einstein-Straus method~\cite{EinStr45} for embedding a point-mass within a homogeneous spacetime (the ``cheese''). It consists in cutting off a spherical domain of the cheese and concentrating the matter it contained at the center of the hole. This section presents the spacetime geometries inside and outside a hole (Subsec.\ref{subsec:spacetime_patches}), and how they are glued together (Subsec.~\ref{subsec:junction_conditions}).

\subsection{Spacetime patches}
\label{subsec:spacetime_patches}

\subsubsection{The ``cheese''---Friedmann-Lema\^itre geometry}

Outside the hole, the geometry is described by the standard Friedmann-Lema\^itre (FL) metric
\begin{equation}\label{eqext}
\dd s^2 =-\dd T^2 + a^2(T)\left[\dd\chi^2 + f_K^2(\chi)\,\dd\Omega^2\right],
\end{equation}
where $a$ is the scale factor and $T$ is the cosmic time. The function $f_K(\chi)$ depends on the sign of $K$ and thus of the spatial geometry (spherical, Euclidean or hyperbolic),
\begin{equation}
f_K(\chi) = \frac{\sin\sqrt{K}\chi}{\sqrt{K}}, \quad \chi \quad \text{or} \quad \frac{\sinh\sqrt{-K}\chi}{\sqrt{-K}}
\end{equation}  
respectively for $K>0$, $K=0$ or $K<0$. The Einstein field equations imply that the scale factor~$a(T)$ satisfies the Friedmann equation
\begin{equation}\label{e.Friedmann}
H^2 = \frac{8\pi G}{3}\rho -\frac{K}{a^2} + \frac{\Lambda}{3},
\quad \text{with} \quad H \equiv \frac{1}{a} \frac{\dd a}{\dd T},
\end{equation}
and where $\rho=\rho_0(a_0/a)^3$ is the energy density of a pressureless fluid. A subscript $0$ indicates that the quantity is evaluated today. It is convenient to introduce the cosmological parameters
\begin{equation}
\Omega_{\rm m}=\frac{8\pi G \rho_0}{3 H_0^2},
\quad
\Omega_K=-\frac{K}{a_0^2H_0^2},
\quad
\Omega_{\Lambda}=\frac{\Lambda}{3 H_0^2},
\end{equation}
in terms of which the Friedmann equation takes the form
\begin{equation}
 \left(\frac{H}{H_0}\right)^2 = \Omega_{\rm m} \left( \frac{a_0}{a} \right)^3 + \Omega_{K} \left( \frac{a_0}{a} \right)^2 +\Omega_{\Lambda} .
\end{equation}

\subsubsection{The ``hole''---Kottler geometry}

Inside the hole, the geometry is described by the extension of the Schwarzschild metric to the case of a nonzero cosmological constant, known as the Kottler solution \cite{Kot18,Wey19} (see e.g. Ref.~\cite{Per04} for a review). In spherical coordinates $(r,\theta,\ph)$, it reads
\begin{align}
\label{eq1c}
\dd s^2 &= -A(r) \, \dd t^2 + A^{-1}(r) \, \dd r^2 + r^2\dd\Omega^2,\\
&\text{with} \quad A(r) \equiv 1 - \frac{r\e{S}}{r} - \frac{\Lambda\,r^2}{3},
\end{align}
and where $r\e{S}\equiv 2\,GM$ is the Schwarzschild radius associated with the mass $M$ at the center of the hole. It is easy to check that the above metric describes a static spacetime. The corresponding Killing vector $\xi^\mu = \delta^\mu_0$ has norm $g_{\mu\nu}\xi^\mu\xi^\nu=A(r)$ and is therefore timelike as long as $A>0$. Hence, there are two cases:
\begin{enumerate}
\item If $9(GM)^2\Lambda>1$, then $A(r)<0$ for all $r>0$, so that $ \xi^\mu$ is spacelike. In this case, the Kottler spacetime contains no static region but it is spatially homogeneous.
\item If $9(GM)^2\Lambda<1$, then $A(r)>0$ for $r$ between $r\e{b}$ and $r\e{c}>r\e{b}$ which are the two positive roots of the polynomial $rA(r)$, and correspond respectively to the black hole and cosmological horizons. We have
\begin{align}
r\e{c} &= \frac{2}{\sqrt{\Lambda}}\cos\left(\frac{\psi}{3} + \frac{\pi}{3} \right),\\
r\e{b} &= \frac{2}{\sqrt{\Lambda}}\cos\left(\frac{\psi}{3} - \frac{\pi}{3} \right),
\end{align}
with $\cos\psi = 3\,GM\sqrt{\Lambda}$, so that
\begin{equation}
r\e{S} < r\e{b} < \frac{3}{2}\,r\e{S} < \frac{1}{\sqrt{\Lambda}} < r\e{c} < \frac{3}{\sqrt{\Lambda}} .
\end{equation}
In the region $r\e{b}<r<r\e{c}$, the Kottler spacetime is static. Note also that $r=r\e{b}$ and $r=r\e{c}$ are Killing horizons, since $\xi$ vanishes on these hypersurfaces. 
\end{enumerate}
In practice, we use the Kottler solution to describe the vicinity of a gravitationally bound object, such as a galaxy, or a cluster of galaxies. In this context, we have typically $9(GM)^2\Lambda < 10^{-14}$ (see Subsec.~\ref{subsec:numerical_values}), so we are in the second case. Moreover, this solution only describes the exterior region of the central object; it is thus valid only for $r>r\e{phys}$, where $r\e{phys}$ is the physical size of the object. For the cases we are interested in, $r\e{phys} \gg r\e{b}$, so that there is actually no black-hole horizon.

\subsection{Junction conditions}
\label{subsec:junction_conditions}

Any spacetime obtained by gluing together two different geometries, via a hypersurface~$\Sigma$, is well defined if---and only if---it satisfies the Israel junction conditions~\cite{Isr66,Isr67}: both geometries must induce (a) the same 3-metric, and (b) the same extrinsic curvature on $\Sigma$.

\begin{figure}[h!]
\centering
\includegraphics[width=\columnwidth]{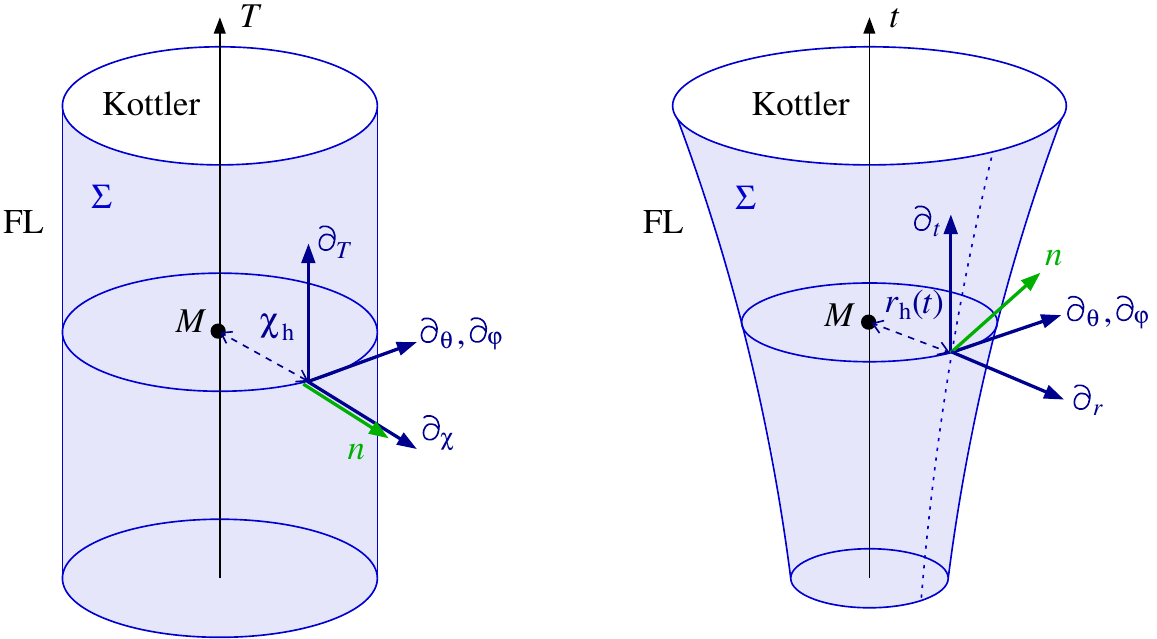}
\caption{The junction hypersurface as seen from the FL point of view with equation $\chi=\chi\e{h}$ (left); and from the Kottler point of view with equation $r=r\e{h}(t)$ (right).}
\label{fig:junction_hypersurface}
\end{figure}

The junction hypersurface $\Sigma$ is the world sheet of a comoving 2-sphere, as imposed by the symmetry of the problem. Hence, it is defined by $\chi=\chi\e{h}=\text{cst}$ in FL coordinates, and by $r=r\e{h}(t)$ in Kottler coordinates. Both points of view are depicted in Fig.~\ref{fig:junction_hypersurface}.

In the FL region, the normal vector to the hypersurface is given by  $n^{({\rm FL})}_\mu=\delta_\mu^\chi/a$. The 3-metric and the extrinsic curvature induced by the FL geometry are respectively
\begin{align}
\dd s^2_\Sigma &= - \dd T^2 + a^2(T)f_K^2(\chi\e{h})\,\dd\Omega^2,\\
K^{({\rm FL})}_{ab}\dd x^a \dd x^b &= a(T)f_K(\chi\e{h})f'_K(\chi\e{h})\,\dd\Omega^2.
\end{align}
where $(x^a) = (T,\theta,\ph)$ are natural intrinsic coordinates for $\Sigma$. We stress carefully that, in the following and as long as there is no ambiguity, a dot can denote a time derivative with respect to $T$ or $t$, so that $\dot a = \dd a/\dd T$ and $\dot r\e{h}=\dd r\e{h}/\dd t$, while a prime can denote a derivative with respect to $\chi$ or $r$, so that $f_K' = \dd f_K/\dd \chi$ and $A'=\dd A/\dd r$.

The 3-metric induced on $\Sigma$ by the Kottler geometry is
\begin{equation}
\dd s^2 = - \kappa^2(t)\,\dd t^2 + r\e{h}^2(t)\,\dd\Omega^2,
\end{equation}
where
\begin{equation}
\kappa(t) \equiv \sqrt{\frac{A^2[r\e{h}(t)]-\dot r\e{h}^2(t)}{A[r\e{h}(t)]}} .
\label{eq:def_kappa}
\end{equation}
Therefore, the first junction condition implies
\begin{align}
r\e{h}(t) &= a(T)f_K(\chi\e{h}),
\label{eq:relation_between_radii}\\
\frac{\dd T}{\dd t} &= \kappa(t),
\label{eq:relation_between_times}
\end{align}
which govern the dynamics of the hole boundary, and relate the time coordinates of the FL and Kottler regions.

The extrinsic curvature of $\Sigma$ induced by the Kottler geometry, but expressed in $(x^a)$ coordinates, reads
\begin{equation}
K^{({\rm K})}_{ab}\dd x^a \dd x^b = - \frac{\ddot r\e{h} + \kappa^2 A'(r\e{h})/2 }{\kappa^3}\,\dd T^2 
									+ \frac{r\e{h} A(r\e{h})}{\kappa} \, \dd\Omega^2 .
\end{equation}
Hence, the second junction condition is satisfied only if
\begin{equation}\label{e.kappa}
\kappa = \frac{A(r\e{h})}{f'_K(\chi\e{h})}, 
\qquad \text{whence} \qquad
\frac{\dd T}{\dd t} = \frac{A[a(T)f_K(\chi\e{h})]}{f'_K(\chi\e{h})} .
\end{equation}
It is straightforward to show that Eq. \eqref{e.kappa}, together with the Friedmann equation \eqref{e.Friedmann}, imply that the Kottler and FL regions have the same cosmological constant, and
\begin{equation}\label{e.massK}
M = \frac{4\pi}{3} \rho \, a^3 f_K^3(\chi\e{h}).
\end{equation}

\subsection{Summary}

Given a FL spacetime with pressureless matter and a cosmological constant, $a(T)$ is completely determined from the Friedmann equation. A spherical hole of comoving radius $\chi\e{h}$, which contains a constant
mass $M =4\pi\rho a^3 f_K^3(\chi\e{h})/3$ at its center, and whose geometry is described by the Kottler metric, can then be inserted anywhere. The resulting spacetime geometry is an exact solution of the Einstein field equations. 

By construction, the clump inside the hole does not backreact on the surrounding FL region. It follows that many such holes can be inserted, as long as they do not overlap. Note that if two holes do not overlap initially, then they will never do so, despite the expansion of the universe, because their boundaries are comoving.

\section{Propagation of light}
\label{sec:geometric_optics}

\subsection{Light rays}

The past light cone of a given observer is a constant phase hypersurface $w =\,$const. Its normal vector $k_\mu \equiv \partial_\mu w$ (the wave four-vector) is a null vector satisfying the geodesic equation, and whose integral curves (light rays) are irrotational:
\begin{equation}\label{nullgeod}
k^\mu k_\mu = 0, \quad k^\nu \nabla_\nu k_\mu = 0, \quad \nabla_{[\mu}k_{\nu]}=0.
\end{equation}
For an emitter and an observer with respective four-velocities $u\e{em}^\mu$ and $u\e{obs}^\mu$, we define the redshift by
\begin{equation}\label{def.z}
1+z =\frac{u\e{em}^\mu k_\mu(v\e{em})}{u\e{obs}^\mu k_\mu(0)},
\end{equation}
where $v$ is an affine parameter along the geodesic, so that $k^\mu=\dd x^\mu/\dd v$, and $v=0$ at the observation event. The wave four-vector can always be decomposed into temporal and spatial components,
\begin{equation}\label{dec-k}
k^\mu =  (1+z)(u^\mu - d^\mu),\quad d^\mu u_\mu =0 ,\quad d^\mu d_\mu=1,
\end{equation}
where $d^\mu $ denotes the spatial direction of observation. In Eq.~\eqref{dec-k}, we have chosen an affine parameter adapted to the observer, in the sense that $2\pi \nu_0 = u\e{obs}^\mu k_\mu(0)=1$. This convention is used in all the remainder of the article.

\subsection{Light beams}

\subsubsection{Geodesic deviation equation}
\label{subsubsec:gde}

A light beam is a collection of light rays, that is, a bundle of null geodesics $\{x^\mu(v,\gamma)\}$, where $\gamma$ labels the curves and $v$ is the affine parameter along them. The relative behavior of two neighboring geodesics $x^\mu(\cdot,\gamma)$ and $x^\mu(\cdot,\gamma+\dd \gamma)$ is described by their separation vector $\xi^\mu \equiv \dd x^\mu/\dd \gamma$. Hence, this vector encodes the whole information on the size and shape of the bundle.

Having chosen $v=0$ at the observation event---which is a vertex point of the bundle---ensures that the separation vector field is everywhere orthogonal to the geodesics, $k^\mu\xi_\mu=0$. In such conditions, the evolution of $\xi^\mu$ with $v$ is governed by the geodesic deviation equation
\begin{equation}\label{gde1}
k^\alpha k^\beta \nabla_\alpha \nabla_\beta \xi^\mu = {R^\mu}_{\nu\alpha\beta} k^\nu k^\alpha\xi^\beta,
\end{equation}
where ${R^\mu}_{\nu\alpha\beta}$ is the Riemann tensor.

\subsubsection{Sachs equation}
\label{subsubsec:Sachs_equation}

Consider an observer with four-velocity $u^\mu$. In view of relating $\xi^\mu$ to observable quantities, we introduce the Sachs basis $(s_A^\mu)_{A\in\{1,2\}}$, defined as an orthonormal basis of the plane orthogonal to both $u^\mu$ and $k^\mu$,
\begin{equation}
s_{A}^\mu s_{B\mu}=\delta_{AB},
\quad
s_{A}^\mu u_{\mu}=s_{A}^\mu k_{\mu}=0,
\end{equation}
and parallel-transported along the geodesic bundle,
\begin{equation}
k^\nu \nabla_\nu s_A^\mu = 0 .
\end{equation}
The plane spanned by $(s_1,s_2)$ can be considered a screen on which the observer projects the light beam. The two-vector of components $\xi_A=\xi_\mu s_A^\mu$ then represents the relative position, on the screen, of the light spots corresponding to two neighboring rays separated by $\xi^\mu$.

The evolution of $\xi_A$, with light propagation, is determined by projecting the geodesic deviation equation \eqref{gde1} on the Sachs basis. The result is known as the Sachs equation~\cite{pubook,sachs,refbooks}, and reads
\begin{equation}
\frac{\dd^2\xi_A}{\dd v^2} = \mathcal{R}_{AB} \, \xi^B,
\label{eq:Sachs}
\end{equation}
where $\mathcal{R}_{AB}={R}_{\mu\nu\alpha\beta}k^\nu k^\alpha s_A^\mu s_B^\beta$ is the screen-projected Riemann tensor, called optical tidal matrix. It is conveniently decomposed into a Ricci term and a Weyl term~as
\begin{equation}\label{eq:decomposition_optical_matrix}
(\mathcal{R}_{AB}) =
\begin{pmatrix}
\Phi_{00} & 0 \\ 0 & \Phi_{00}
\end{pmatrix}
+
\begin{pmatrix}
- {\rm Re}\,\Psi_0 & {\rm Im}\,\Psi_0 \\
  {\rm Im}\,\Psi_0 & {\rm Re}\,\Psi_0
\end{pmatrix}
\end{equation}
with
\begin{equation}\label{ricweyl}
 \Phi_{00} \equiv -\frac12 R_{\mu\nu} k^\mu k^\nu,\quad
 \Psi_0 \equiv -\frac{1}{2} C_{\mu\nu\alpha\beta} \, \sigma^\mu
 k^\nu k^\alpha \sigma^\beta,
\end{equation}
and where $\sigma^\mu \equiv s_1^{\mu} - {\rm i} \, s_2^{\mu}$.

\subsubsection{Notions of distance}

Since the light beam converges at the observation event, we have $\xi^A(v=0)=0$. The linearity of the Sachs equation then implies the existence of a $2\times 2$ matrix ${\cal D}\indices{^A_B}$, called Jacobi matrix, such that
\begin{equation}\label{jacobimat}
\xi^A(v) = {\cal D}\indices{^A_B}(v) \left(\frac{\dd\xi^B}{\dd v}\right)_{v=0}.
\end{equation}
From Eq.~\eqref{eq:Sachs}, we immediately deduce that this matrix satisfies the Jacobi matrix equation
\begin{equation}\label{gde2}
\frac{\dd^2}{\dd v^2}\,{\cal D}\indices{^A_B}={\cal R}\indices{^A_C} \, {\cal D}\indices{^C_B},
\end{equation}
with initial conditions
\begin{equation}\label{gde2b}
{\cal D}\indices{^A_B}(0) =0,
\qquad
\frac{\dd{\cal D}\indices{^A_B}}{\dd v}(0)=\delta^A_B.
\end{equation}
We shall also use the short-hand notation $\boldsymbol{\xi}=(\xi^A)$ and $\boldsymbol{\mathcal D}=({\cal D}\indices{^A_B})$ so that Eq.~(\ref{gde2}) reads $\dd^2\boldsymbol{\mathcal D}/\dd v^2=\boldsymbol{\mathcal R}\cdot\boldsymbol{\mathcal D}$, with $\boldsymbol{\mathcal D}(0)=0$ and $\boldsymbol{\dot{\mathcal D}}(0)=\boldsymbol{1}$.

Since the Jacobi matrix relates the shape of a light beam to its ``initial'' aperture, it is naturally related to the various notions of distance used in astronomy and cosmology. The \emph{angular distance} $D\e{A}$ is defined by comparing the emission cross-sectional area $\dd^2 S_{\rm source}$ of a source to the solid angle $\dd\Omega_{\rm obs}^2$ under which it is observed,
\begin{equation}
\dd^2 S_{\rm source}=D\e{A}^2 \, \dd\Omega_{\rm obs}^2.
\end{equation}
It is related to the Jacobi matrix by
\begin{equation}\label{defDA}
D\e{A} = \sqrt{\vert{\rm det}\,{\boldsymbol{\mathcal D}}(v_{\rm source})\vert},
\end{equation}
where $v\e{source}$ is the affine parameter at emission.

The \emph{luminosity distance} $D\e{L}$ is defined from the ratio between the observed flux~$F\e{obs}$ and the intrinsic luminosity~$L\e{source}$ of the source, so that
\begin{equation}
L\e{source} = 4\pi D\e{L}^2 \, F\e{obs} .
\end{equation}
It is related to the angular distance by the following distance duality law
\begin{equation}
D\e{L}=(1+z)^2 D\e{A}.
\end{equation}
Hence, the theoretical determination of the luminosity distance relies on the computation of the Jacobi matrix.

\subsection{Solving the Sachs equation piecewise}
\label{sec-piece}

Since we work in a Swiss-cheese universe, we have to compute the Jacobi matrix for a patchwork of spacetimes. It is tempting, in this context, to calculate the Jacobi matrix for each patch independently, and then try to reconnect them. In fact, such an operation is unnatural, because the very definition of $\boldsymbol{\mathcal{D}}$ imposes that the initial condition is a vertex point of the light beam. Thus, juxtaposing two Jacobi matrices is only possible at a vertex point, which is of course too restrictive for us.

We can solve this problem by extending the Jacobi matrix formalism into a richer structure. This requires us to consider the general solution of Eq.~\eqref{eq:Sachs}, for arbitrary initial conditions. Thus, we have
\begin{equation}\label{jacobimat2}
\boldsymbol{\mathcal \xi}(v) = \boldsymbol{\mathcal C}(v;v\e{init})\cdot \boldsymbol{\mathcal \xi}_{v=v\e{init}} +
\boldsymbol{\mathcal D}(v;v\e{init})\cdot \left.\frac{\dd\boldsymbol{\mathcal \xi}}{\dd v}\right\vert_{v=v\e{init}} ,
\end{equation} 
as for any linear second order differential equation, solved from $v\e{init}$ to $v$. In the following, $\boldsymbol{\mathcal C}(v;v\e{init})$ is referred to as the {\em scale matrix}. It is easy to check that both the scale and Jacobi matrices satisfy the Jacobi matrix equation~(\ref{gde2}) but with different initial conditions:
\begin{equation}\label{gde2c}
\boldsymbol{\mathcal D}(v\e{init};v\e{init}) = \boldsymbol{0}, 
\qquad
\frac{\dd\boldsymbol{\mathcal D}}{\dd v}(v\e{init};v\e{init}) = \boldsymbol{1},
\end{equation}
whereas
\begin{equation}\label{gde2d}
\boldsymbol{\mathcal C}(v\e{init};v\e{init}) = \boldsymbol{1}, 
\qquad
\frac{\dd\boldsymbol{\mathcal C}}{\dd v}(v\e{init};v\e{init}) = \boldsymbol{0}.
\end{equation}

The most useful object for our problem turns out to be the $ 4\times 4$ \emph{Wronski matrix} constructed from $\boldsymbol{\mathcal C}$ and $\boldsymbol{\mathcal D}$,
\begin{equation}
 \boldsymbol{\mathcal W}(v;v\e{init})\equiv\left(
 \begin{array}{cc}
  \boldsymbol{\mathcal C}(v;v\e{init}) & \boldsymbol{\mathcal D}(v;v\e{init})\\
  \frac{\dd\boldsymbol{\mathcal C}}{\dd v}(v;v\e{init}) & \frac{\dd\boldsymbol{\mathcal D}}{\dd v}(v;v\e{init})
 \end{array}
 \right),
\end{equation}
in terms of which the general solution~(\ref{jacobimat2}) reads
\begin{equation}\label{eq:use_wronski}
\left(
 \begin{array}{c}
  \boldsymbol{\xi}\\
  \frac{\dd\boldsymbol{\xi}}{\dd v}
 \end{array}
 \right)(v)=  \boldsymbol{\mathcal W}(v;v\e{init})\cdot
 \left(
 \begin{array}{c}
  \boldsymbol{\xi}\\
  \frac{\dd\boldsymbol{\xi}}{\dd v}
 \end{array}
 \right)(v\e{init}).
\end{equation}
It is clear, from Eq.~\eqref{eq:use_wronski}, that $\boldsymbol{\mathcal W}$ satisfies the relation
\begin{equation}
  \boldsymbol{\mathcal W}(v_1;v_3)= \boldsymbol{\mathcal W}(v_1;v_2) \cdot \boldsymbol{\mathcal W}(v_2;v_3) .
\end{equation}
Hence, the general solution of the Sachs equation in a Swiss-cheese universe can be obtained by multiplying Wronski matrices, according to
\begin{multline}\label{e.chain}
  \boldsymbol{\mathcal W}(v_{\rm source};0) = \boldsymbol{\mathcal W}_{\rm FL}(v_{\rm source};v_{\rm in}^{(1)})\cdot \boldsymbol{\mathcal W}_{\rm K}(v_{\rm in}^{(1)};v_{\rm out}^{(1)}) \\
 \cdot \boldsymbol{\mathcal W}_{\rm FL}(v_{\rm out}^{(1)};v_{\rm in}^{(2)})\cdots \boldsymbol{\mathcal W}_{\rm FL}(v_{\rm out}^{(N)};0)
\end{multline}
where $ \boldsymbol{\mathcal W}_{\rm FL}$ and  $\boldsymbol{\mathcal W}_{\rm K}$ are the Wronski matrices computed respectively in the FL region and in the Kottler holes; $v_{\rm in}^{(i)}$ and $v_{\rm out}^{(i)}$ are the values of the affine parameter respectively at the entrance and the exit of the $i$th hole.

\section{Integration of the geodesic and Sachs equations}
\label{sec:resolution}

Consider an observer lying within a FL region, who receives a photon after the latter has crossed a hole. In this section, we determine the light path from entrance to observation by solving the geodesic equation, and we calculate the Wronski matrix for the Sachs equation.

\begin{widetext}

\begin{figure}[htb]
\centering
\includegraphics[scale=1]{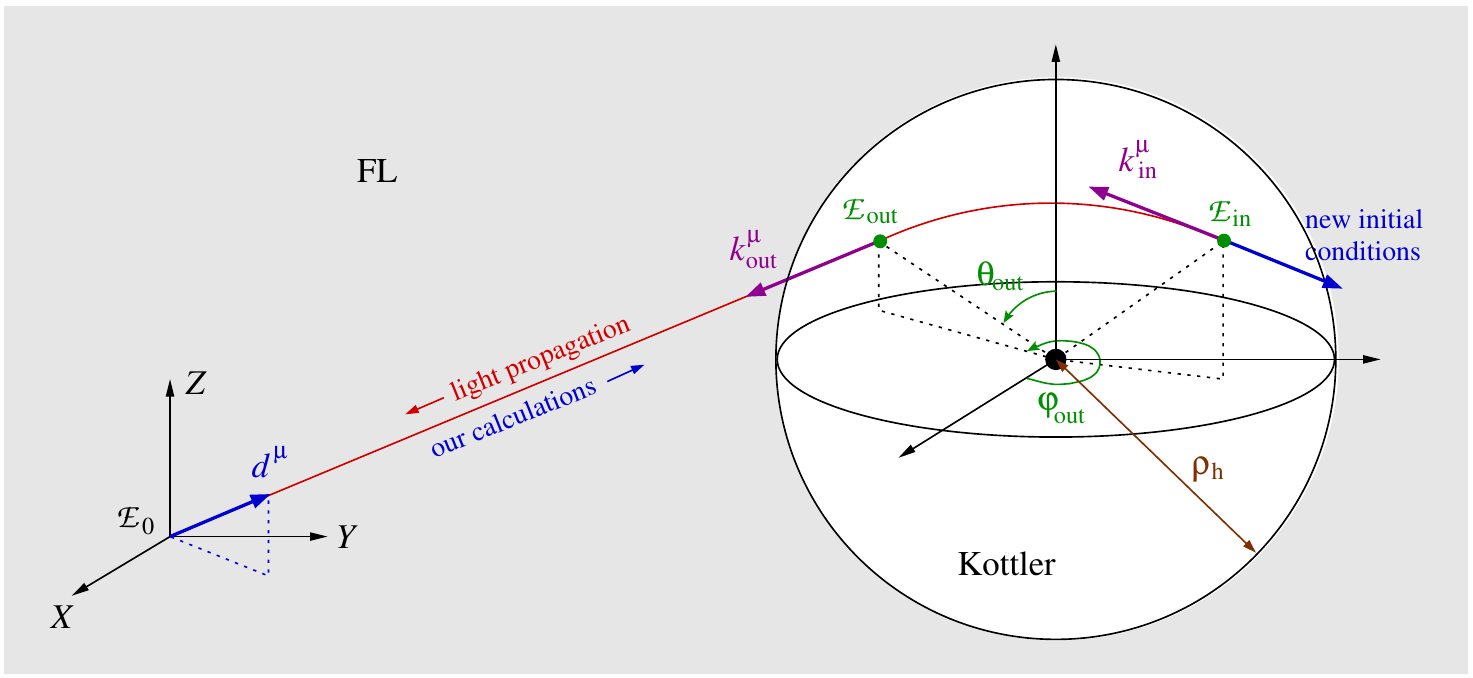}
\caption{A light ray propagates alternatively in FL and Kottler regions. The main geometrical quantities defined and used in Sec.~\ref{sec:resolution} are depicted in this simplified view of a single hole.}
\label{figgeo}
\end{figure}

\end{widetext}

The main geometrical quantities are summarized in Fig.~\ref{figgeo}. $d^\mu$ is the direction of observation as defined in Eq.~(\ref{dec-k}). The spatial sections of the FL region can be described either by comoving spherical coordinates $(\chi,\theta,\varphi)$ or, when the spatial sections are Euclidean, by comoving Cartesian coordinates $(X,Y,Z)$. 

The hole is characterized by its comoving spatial position~$X_{\rm h}^i$, in terms of the FL coordinates, and its mass~$M$, or equivalently its comoving radius~$\rho_{\rm h}$. Note that, contrary to Subsec.~\ref{subsec:spacetime_patches}, it is no longer denoted~$\chi\e{h}$, in order to avoid confusion with the radial comoving coordinate of the center of the hole.

A photon enters into the hole with wave vector $k^\mu_{\rm in}$, exits from it with wave vector $k^\mu_{\rm out}$, and reaches the observer with wave vector $k^\mu_0$. We respectively denote ${\cal E}_{\rm in}$, ${\cal E}_{\rm out}$ and ${\cal E}_0$ the associated events. The coordinates of the first two can be expressed either with respect to FL, e.g. as $(T_{\rm in},X^i_{\rm in})$ in Cartesian coordinates, or with respect to the hole, e.g. as $(t_{\rm in},r_{\rm in},\theta_{\rm in},\varphi_{\rm in})$ in the Kottler spherical coordinate system.

Our calculations go backward in time. Starting from $\mathcal{E}_0$, we first determine $\mathcal{E}\e{out}$, $\boldsymbol{\mathcal{W}}_{\rm FL}(v_{\rm out};v\e{obs})$, and second $\mathcal{E}\e{in}$, $\boldsymbol{\mathcal{W}}_{\rm K}(v_{\rm in};v\e{out})$. The same operations can then be repeated starting from $\mathcal{E}\e{in}$ and so on.

\subsection{Friedmann-Lema\^{\i}tre region {\it(from ${\cal E}_0$ to ${\cal E}_{\rm out}$)}}
\label{secW-FL} 

The geometry of the Friedmann-Lema\^{\i}tre region is given by the metric~(\ref{eqext}) which can be rewritten in terms of the conformal time $\eta$, defined by $\dd\eta =\dd T/a(T)$, as
\begin{equation}\label{e-fleq}
\dd s^2 = a^2(\eta)\left[-\dd\eta^2 + \dd\chi^2 + f_K^2(\chi)\,\dd\Omega^2\right].
\end{equation}

\subsubsection{Geodesic equation}

If one chooses the center $\chi=0$ of the FL spherical coordinate system on the worldline of the (comoving) observer, then the geodesic equation is easily solved as
\begin{equation}\label{e.FLradial}
 \chi(\eta) = \eta_0 - \eta, \quad
 \theta=\theta_0,\quad
 \varphi=\varphi_0,
\end{equation}
which corresponds to a purely radial trajectory. Note however that for a generic origin, this is no longer true. The associated wave vector remains collinear to the observed one,~$k^\mu_0$. It is only subject to a redshift induced by the cosmic expansion, so that 
\begin{equation}
k^\mu = \left(\frac{a_0}{a}\right)^2 k^\mu_0 .
\label{e.wavevectorFL}
\end{equation}
We stress that, in Eq.~\eqref{e.wavevectorFL}, $\mu=0$ refers to components on $\partial_\eta$, not on $\partial_T=\partial_\eta/a$.

\subsubsection{Intersection with the hole}

Once the geodesic equation has been solved and the position of the hole has been chosen, we can calculate the intersection ${\cal E}_{\rm out}$ between the light ray and the hole boundary. \emph{In the particular case of a spatially Euclidean FL solution} ($K=0$), the Cartesian coordinates $X^ i\e{out}$ of ${\cal E}_{\rm out}$ satisfy the simple system of equations
\begin{equation}
\left\{
\begin{aligned}
& \delta_{ij}\left(X^i_{\rm out} - X_{\rm h}^i\right)\left(X^j_{\rm out} - X_{\rm h}^j\right)=\rho_{\rm h}^2\\
& X^i_{\rm out} = X^i_0 +(\eta_0-\eta_{\rm out})\,d^i
\end{aligned}
\right.
,
\end{equation}
where $X_{\rm h}^i$ and $X^i_0$ are the respective Cartesian coordinates of the hole and the observer, while $d^i$ is the spatial direction of observation. Although conceptually similar, the determination of $\mathcal{E}\e{out}$ for a FL solution with arbitrary spatial curvature is technically harder.

In general, we deduce from Eq.~\eqref{e.wavevectorFL} that the wave vector at $\mathcal{E}\e{out}$ is $k^\mu\e{out} = (a_0/a\e{out})^ 2 k^\mu_0$, where $a\e{out}\equiv a(\eta\e{out})$.

\subsubsection{Wronski matrix}

In the FL region, the Sachs basis~$(s_1,s_2)$ is defined with respect to the fundamental observers, comoving with four-velocity $u=\partial_T$. The explicit form of this basis does not need to be specified here.

The Sachs equation can be solved analytically by means of a conformal transformation to the static metric
\begin{equation}\label{e.conf}
\dd \tilde s^2 = a_0^2\left[-\dd\eta^2 + f_K^2(\chi)\dd\Omega^2\right]\equiv\tilde g_{\mu\nu}\dd x^\mu\dd x^\nu.
\end{equation}
Because the geometries associated with $g_{\mu\nu}$ and $\tilde g_{\mu\nu}$ are conformal, any null geodesic for $g_{\mu\nu}$ affinely parametrized by $v$ is also a null geodesic for $\tilde g_{\mu\nu}$ affinely parametrized by $\tilde{v}$, with $a^2  \dd \tilde{v} = a_0^2 \dd v$. As $\dd v = (a^2/a_0)\dd\eta$, it follows that $\tilde{v}=a_0\eta$.

For the static geometry, the optical tidal matrix reads $\widetilde{\boldsymbol{\mathcal R}}=-(K/a_0^2) \boldsymbol{1}$, so that the Sachs equation is simply
\begin{equation}\label{gdefl}
\frac{\dd^2\tilde{\boldsymbol{\mathcal \xi}}}{\dd \eta^2}= - K \,\tilde{\boldsymbol{\mathcal \xi}}.
\end{equation}
We then easily obtain the Jacobi and scale matrices:
\begin{equation}
 \widetilde{\boldsymbol{\mathcal D}} = a_0 f_K(\eta-\eta\e{init})\boldsymbol{1},\qquad
 \widetilde{\boldsymbol{\mathcal C}} = f'_K(\eta-\eta\e{init})\boldsymbol{1}.
\end{equation}

To go back to the original FL spacetime, we use that $\dd v= a^2\dd \eta$ and the fact that the screen projections of the separation vectors for both geometries are related by $a \tilde{\boldsymbol{\xi}}=a_0 \boldsymbol{\xi}$. The final result is
\begin{align}
{\boldsymbol{\mathcal D}}\e{FL} &= a\e{init }\frac{a}{a_0} f_K(\eta-\eta\e{init}) \boldsymbol{1},\\
{\boldsymbol{\mathcal C}}\e{FL} &= \frac{a}{a\e{init}} \Big[ f'_K(\eta-\eta\e{init}) - \mathcal{H}\e{init} f_K(\eta-\eta\e{init}) \Big] \boldsymbol{1},
\end{align}
where $\mathcal{H} \equiv a'(\eta)/a(\eta)$ is the conformal Hubble function. This completely determines ${\boldsymbol{\mathcal W}}\e{FL}$.

Note that we can recover the standard expression of the angular distance by taking the initial condition at the observer. The relation~(\ref{defDA}) then implies
\begin{equation}
D\e{A} = \sqrt{\det\boldsymbol{\mathcal{D}\e{FL}}} = \frac{a_0}{(1+z)} \, f_K(\eta\e{source}),
\label{eq:FL_DA-z_relation}
\end{equation}
where $z=a_0/a-1$ is the redshift of a photon that only travels through a FL region.

\subsection{Kottler region {\it (from ${\cal E}_{\rm out}$ to ${\cal E}_{\rm in}$)}}
\label{subsec:Kottler}

\subsubsection{Initial condition at ${\cal E}_{\rm out}$}
\label{subsubsec:out_conditions}

In the previous section, we have determined ${\cal E}_{\rm out}$ and $k^\mu_{\rm out}$ in terms of the FL coordinate system. However, in order to proceed inside the hole, we need to express them in terms of the Kottler coordinate system $(t,r,\theta,\ph)$.

A preliminary task consists in expressing ${\cal E}_{\rm out}$ and $k^\mu_{\rm out}$ in terms of FL spherical coordinates, with origin at the center of the hole. This operation is straightforward. The event ${\cal E}_{\rm out}$ is then easily converted, since (a) we are free to set $t\e{out}=0$, (b) Eq.~\eqref{eq:relation_between_radii} implies $r\e{out}=a(\eta\e{out})\rho\e{h}$, and (c) the angular coordinates $\theta\e{out}$, $\ph\e{out}$ remain unchanged if the Kottler axes are chosen parallel to the FL ones.

The first junction condition ensures that light is not deflected when it crosses the boundary~$\Sigma$ of the hole. Indeed, the continuity of the metric implies that the connection does not diverge on~$\Sigma$. Integrating the geodesic equation $\dd k^\mu=-\Gamma^\mu_{\alpha \beta} k^\alpha k^\beta \dd v$ between $v\e{out}^-$ and $v\e{out}^+$ then shows that $k^\mu$ is continuous at $\mathcal{E}\e{out}$. Therefore, we just need to convert its components from the FL coordinate system to the Kottler one. The result is
\begin{align}
  k^t_{\rm out} &= \frac{a_{\rm out}}{A(r_{\rm out})} \left[k_{\rm out}^\eta + \sqrt{1-A(r_{\rm out})} k_{\rm out}^\chi \right]
  \label{eq:ktout}
  \\
  k^r_{\rm out} &= a_{\rm out}\left[\sqrt{1-A(r_{\rm out})} k_{\rm out}^\eta  +  k_{\rm out}^\chi \right]
  \\
  k^\theta_{\rm out} &= k^\theta_{\rm out} 
  \\
  k^\varphi_{\rm out} &= k^\varphi_{\rm out}.
\end{align}

\subsubsection{Shifting to the equatorial plane}
\label{subsubsec:rotations}

Since the Kottler spacetime is spherically symmetric, it is easier to integrate the geodesic equation in the equatorial plane $\theta=\pi/2$. In general, however, we must perform rotations to bring both ${\cal E}_{\rm out}$ and $k_{\rm out}$ into this plane.

\begin{widetext}

\begin{figure}[h!]
\centering
\includegraphics[width=5cm]{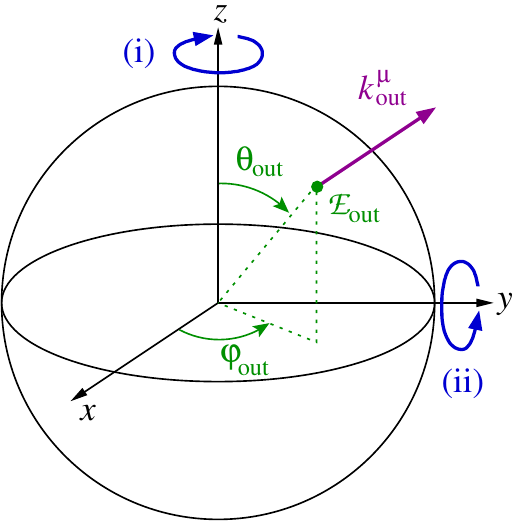}
\qquad
\includegraphics[width=5cm]{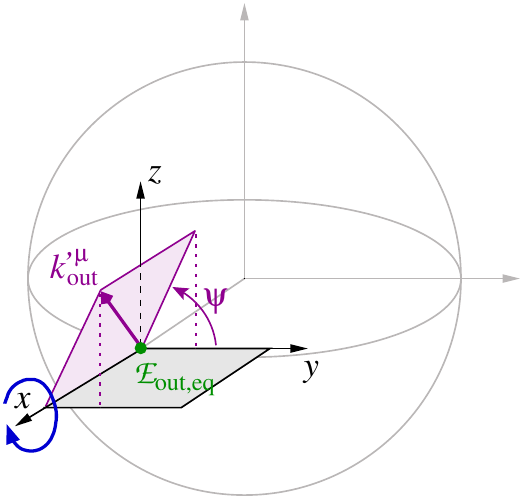}
\qquad
\includegraphics[width=5cm]{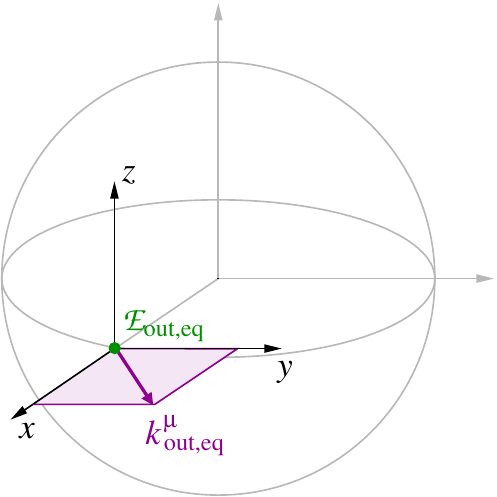}
\caption{An arbitrary initial condition is rotated so that the geodesic lies in the equatorial plane $\theta=\pi/2$. Left: $\mathcal{E}\e{out}$ is brought (i) to $\ph=0$ by the rotation $\boldsymbol{R}_z(-\varphi_{\rm out})$, and (ii) to $\theta=\pi/2$ by the rotation $\boldsymbol{R}_y(\pi/2-\theta_{\rm out})$. The resulting event and wave vector are denoted ${\cal E}_{\rm out,eq}$ and $k'^\mu_{\rm out}$. Middle: $k'^\mu\e{out}$ is brought to the equatorial plane by the rotation $\boldsymbol{R}_x(-\psi)$. Right: Final situation.}
\label{fig:rotations}
\end{figure}

\end{widetext}

Starting from arbitrary initial conditions $({\cal E}_{\rm out}, k^\mu_{\rm out})$, we can shift to the equatorial plane in two steps. In the following, $\boldsymbol{R}_i(\vartheta)$ denotes the rotation of angle $\vartheta$ about the $x^i$-axis. The operations are depicted in Fig.~\ref{fig:rotations}.
\begin{itemize}
\item First, bring ${\cal E}_{\rm out}$ to the point ${\cal E}_{\rm out, eq}$ on the equatorial plane by the action of two successive rotations, $\boldsymbol{R}_z(-\varphi_{\rm out})$ followed by $\boldsymbol{R}_y(\pi/2-\theta_{\rm out})$. The wave vector after the two rotations is denoted $k^{\prime\mu}_{\rm out}$.
\item Then, bring $k^{\prime\mu}_{\rm out}$ to the equatorial plane with $\boldsymbol{R}_x(-\psi)$, where $\psi$ is the angle between the projection of $k^{\prime\mu}_{\rm out}$ on the $yz$-plane and the $y$-axis. Note that such a rotation leaves ${\cal E}_{\rm out, eq}$ unchanged.
\end{itemize}

It follows that, after the three rotations
\begin{equation}
\boldsymbol{R} =
\boldsymbol{R}_x(-\psi)
\circ
\boldsymbol{R}_y\left(\frac{\pi}{2}-\theta_{\rm out}\right) 
\circ 
\boldsymbol{R}_z(-\varphi_{\rm out}),
\end{equation}
${\cal E}_{\rm out}$ and $k^\mu_{\rm out}$ are changed into ${\cal E}_{\rm out,eq}$ and $k^\mu_{\rm out,eq}$ which lie in the equatorial plane. In the following, we omit subscripts ``eq,'' keeping in mind that we will have to apply~$\boldsymbol{R}^{-1}$ to recover the original system of axes.

\subsubsection{Null geodesics in Kottler geometry\protect\footnote{See e.g. Refs.~\cite{Stuchlik1983,Stuchlik1984} for early works on the propagation of light rays in spacetimes with a non-vanishing cosmological constant.}}

In the Kottler region, the existence of two Killing vectors associated to statisticity and spherical symmetry implies the existence of two conserved quantities, the energy $E$ and the angular momentum $L$ of the photon. It follows that a null geodesic is a solution of
\begin{equation}
A(r)\frac{\dd t}{\dd v} = E,
\quad
\left(\frac{\dd r}{\dd v}\right)^2 + \left(\frac{L}{r}\right)^2 A(r) = E^2,
\quad
r^2\frac{\dd \varphi}{\dd v} = L.
\label{eq:Kottler_geodesic_bare}
\end{equation}
Introducing the dimensionless variable $u\equiv r\e{S}/r$ and the impact parameter $b=L/E$, Eqs.~\eqref{eq:Kottler_geodesic_bare} imply
\begin{align}
r\e{S}^2 \left(\frac{\dd u}{\dd t}\right)^2 
&= \frac{u^4}{\varepsilon_1^2} \, P(u) \, A^2(u),
\label{e.tdeuK}\\
\left(\frac{\dd u}{\dd\varphi}\right)^2 
&= P(u),
\label{e.udephiK}\\
\frac{r\e{S}^2}{E^2} \left(\frac{\dd u}{\dd v}\right)^2 
&= \frac{u^4}{\varepsilon_1^2} \, P(u),
\label{e.udev}
\end{align}
with
\begin{equation}
A(u) = 1 - u - \varepsilon_2 u^{-2}, \quad  P(u) \equiv \varepsilon_1^2 - u^2 A(u),
\label{eq:def_P}
\end{equation}
and where $\varepsilon_1\equiv r\e{S}/b$ and $\varepsilon_2\equiv\Lambda r\e{S}^2/3$.

\begin{figure}[h!]
\centering
\includegraphics[width=0.75\columnwidth]{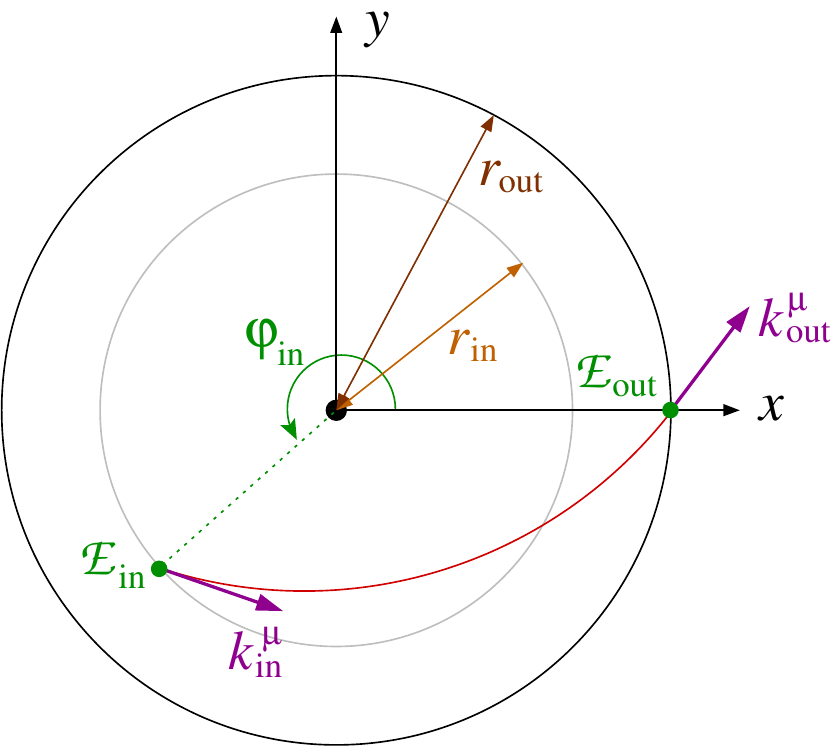}
\caption{Null geodesic in the Kottler region. Depicted with the Kottler coordinate system, the hole grows so that the ray enters with $r_{\rm in}$ and exits with $r_{\rm out}>r_{\rm in}$.}
\label{fig:deflection}
\end{figure}

Our purpose is now to compute the coordinates $(t\e{in},r\e{in},\ph\e{in})$ and the components $k\e{in}^\mu$ of the wave vector at the entrance event $\mathcal{E}\e{in}$, given those at $\mathcal{E}\e{out}$. The situation is summarized in Fig. \ref{fig:deflection}.

The radius $r\e{in}$ (or alternatively $u\e{in}$) and time $t\e{in}$ at entrance are determined by comparing the radial dynamics of the photon, governed by Eq. \eqref{e.tdeuK}, to the one of the hole boundary. The latter is obtained from Eqs. \eqref{eq:def_kappa} and \eqref{e.kappa}. By introducing $u\e{h}=r\e{S}/r\e{h}$, it reads
\begin{equation}
r\e{S}\frac{\dd u_{\rm h}}{\dd t} =- u_{\rm h}^2A(u_{\rm h})\sqrt{1-A(u_{\rm h})} .
\label{eq:dynamics_hole}
\end{equation}
Equations \eqref{e.tdeuK} and \eqref{eq:dynamics_hole} are then integrated\footnote{The integration can be performed either numerically, or analytically in the case of Eq.~\eqref{e.tdeuK} and perturbatively for Eq.~\eqref{eq:dynamics_hole}.} as $t\e{photon}(u)$ and $t\e{hole}(u\e{h})$. The entrance radius then results from solving numerically the equation $t\e{photon}(u\e{in})=t\e{hole}(u\e{in})$, which also provides $t\e{in}$.

The usual textbook calculation of the deflection angle $\Delta\ph_\infty$ of a light ray in Kottler geometry yields
\begin{equation}\label{e.dfS}
\Delta\varphi_\infty = 2 \, \eps_1 \sqrt{1 + \frac{\eps_2}{\eps_1^2}} = \frac{4\,GM}{b} \sqrt{ 1 + \frac{\Lambda\,b^2}{3} }
\end{equation} 
at lowest order in $\eps_1$ and $\eps_2$. However, we cannot use this expression here---although it gives its typical order of magnitude---because $\Delta\varphi_\infty$ represents the angle between the \emph{asymptotic} incoming and outgoing directions of a ray, whereas we must take into account the \emph{finite} extension of the hole (see Fig. \ref{fig:deflection}).

In general, the deflection angle $\Delta\ph=\ph\e{out}-\ph\e{in}$ is
\begin{equation}
\Delta\varphi =\int_{u_{\rm in}}^{u_{\rm m}}\frac{\dd u}{\sqrt{P(u)}} + \int_{u_{\rm out}}^{u_{\rm m}}\frac{\dd u}{\sqrt{P(u)}} -2\pi
\label{eq:deflection_angle}
\end{equation}
where $P(u)$ is the polynomial defined in Eq. \eqref{eq:def_P}, and $u_{\rm m}$ is the value of $u$ at minimal approach. The integral involved in Eq.~\eqref{eq:deflection_angle} can be rewritten as
\begin{equation}\label{eq:integral_and_elliptic}
\int_{u}^{u\e{m}}\frac{\dd u'}{\sqrt{P(u')}} = \frac{2}{\sqrt{u_3-u_2}} \,
											F\left[\arcsin\sqrt{\frac{u_2-u}{u_2-u_1}},\frac{u_2-u_1}{u_2-u_3}\right],
\end{equation}
where $u_1<u_2=u_{\rm m}<u_3$ are the three (real) roots of $P(u)$, and $F(\psi,e)$ denotes the elliptic function of the first kind \cite{GradshteynRyzhik}
\begin{equation}
F(\psi,e) \equiv \int_0^\psi \frac{\dd\theta}{ \sqrt{1 - e \sin^2\theta} } .
\end{equation}
Thus, Eq.~\eqref{eq:integral_and_elliptic} provides an exact expression of the deflection angle~$\Delta\ph$, and therefore of $\ph\e{in}$.

Once ${\cal E}_{\rm in}$ is determined, it is easy to obtain $k^\mu_{\rm in}$ by using the constants of motion. The result is
\begin{align}
k_{\rm in}^t &= \frac{E}{A(r_{\rm in})} 
= \frac{A(r_{\rm out})}{A(r_{\rm in})} \, k^t_{\rm out}, \\
k_{\rm in}^\varphi &= \frac{L}{r_{\rm in}^2}
= \left(\frac{r_{\rm out}}{r_{\rm in}}\right)^2 
	k^\varphi_{\rm out}, \\
k_{\rm in}^r &= -\sqrt{\left[A(r_{\rm in})\,k_{\rm in}^t \right]^2 
- A(r_{\rm in})\left(r_{\rm in}\,k_{\rm in}^\varphi  \right)^2}.
\end{align}

\subsubsection{Final conditions at ${\cal E}_{\rm in}$}

The last step consists in coming back to the original FL coordinate system. That means (a) using $\boldsymbol{R}^{-1}$ to recover the initial system of axes, and (b) converting the components of $\mathcal{E}\e{in}$ and $k\e{in}^\mu$ in terms of the FL coordinate system. We have already described such operations in \S~\ref{subsubsec:rotations} and \S~\ref{subsubsec:out_conditions} respectively, except for the time coordinate (since we set $t\e{out}=0$).

The easiest way to compute the cosmic time~$T\e{in}$ at entrance is to use the relation $r\e{in}=a(T\e{in}) f_K(\rho\e{h})$. In a spatially Euclidean FL spacetime ($K=0$), we get
\begin{equation}
T\e{in} = \frac{2}{3 H_0 \sqrt{\Omega_{\Lambda}} } \,
			\mathrm{argsinh}\left[ \sqrt{\frac{\Omega_{\Lambda}}
									{1-\Omega_{\Lambda}}}
							\left( \frac{r\e{in}}{a_0 \rho\e{h}} \right)^{3/2}
					\right] .
\end{equation}
With this last result, we have completely determined the entrance event $\mathcal{E}\e{in}$.

\subsubsection{Sachs basis and optical tidal matrix}

Once the geodesic equation is completely solved, we are ready to integrate the Sachs equation in the Kottler region, that is, to determine the Wronski matrix~$\boldsymbol{\mathcal{W}}\e{K}$. Such a task requires us first to define the Sachs basis $(s_1,s_2)$ with respect to which $\boldsymbol{\mathcal{W}}\e{K}$ will be calculated.

The four-velocity $u$ is chosen to be the one of a radially free-falling observer,
\begin{equation}
u \equiv \frac{1}{A(r)} \, \partial_t + \sqrt{1-A(r)} \, \partial_r .
\label{eq:Sachs_basis_Kottler_u}
\end{equation}
This choice ensures the continuity of $u$ through the hole frontier, where $u=\partial_T$. The wave four-vector $k$ is imposed by the null geodesic equations, and reads
\begin{equation}
k = \frac{E}{A(r)} \, \partial_t 
			\pm E \sqrt{1-\frac{b^2 A(r)}{r^2}} \, \partial_r 
			+ \frac{L}{r^2} \, \partial_\ph
\label{eq:Sachs_basis_Kottler_k}
\end{equation}
where the $\pm$ sign depends on whether the photon approaches ($-$) or recedes ($+$) from the center of the hole.

By definition, the screen vectors $s_1$, $s_2$ form an orthonormal basis of the plane orthogonal to both $u$ and $k$. Here, since the trajectory occurs in the equatorial plane, the first one can be trivially chosen as
\begin{equation}
s_1 \equiv \partial_z = -\frac{1}{r} \, \partial_\theta .
\label{eq:Sachs_basis_Kottler_s1}
\end{equation}
The second one is obtained from the orthogonality and normalization constraints, and reads
\begin{multline}
s_2 
\equiv 
\frac{1}{N} \bigg[ \frac{\sqrt{1-A(r)}}{A(r)}\,\partial_t 
					+ \partial_r \\
					+ \frac{1}{b\,A(r)} 
						\left(
						\sqrt{1-A(r)}
						\mp \sqrt{1-\frac{b^2 A(r)}{r^2}}
						\right)
						\partial_\ph 	
			\bigg] ,
\label{eq:Sachs_basis_Kottler_s2}
\end{multline}
where the normalization function is
\begin{equation}
N \equiv \frac{r}{b\,A(r)} 
\left( 
1 \mp \sqrt{1-A(r)} \sqrt{1-\frac{b^2 A(r)}{r^2}}
\right) .
\end{equation}
Using the Sachs basis defined by Eqs.~\eqref{eq:Sachs_basis_Kottler_u},  \eqref{eq:Sachs_basis_Kottler_k}, 
\eqref{eq:Sachs_basis_Kottler_s1}, and \eqref{eq:Sachs_basis_Kottler_s2}, we can finally compute the optical tidal matrix, and get
\begin{equation}
\boldsymbol{\mathcal{R}} =
\begin{pmatrix}
-\mathcal{R}(r) & 0 \\
0 & \mathcal{R}(r)
\end{pmatrix} ,
\end{equation}
where the function $\mathcal{R}(r)$ is
\begin{equation}
\mathcal{R}(r) \equiv \frac{3}{2} \left( \frac{L}{r\e{S}^2} \right)^2 									\left( \frac{r\e{S}}{r} \right)^5 .
\label{eq:projected_Riemann_tensor_Schwarzschild}
\end{equation}
As expected from the general decomposition \eqref{eq:decomposition_optical_matrix}, ${\boldsymbol{\mathcal R}}$ is trace free because only Weyl focusing is at work. Let us finally emphasize that $\Lambda$ does not appear in the expression \eqref{eq:projected_Riemann_tensor_Schwarzschild} of $\mathcal{R}(r)$, which is not surprising since a pure cosmological constant does not deflect light.

\subsubsection{Wronski matrix}
\label{paragraph:perturbative_Wronski}

The Sachs equations can now be integrated in order to determine the scale matrix~$\boldsymbol{\mathcal{C}}\e{K}$ and the Jacobi matrix~$\boldsymbol{\mathcal{D}}\e{K}$ that compose the Wronski matrix~$\boldsymbol{\mathcal{W}}\e{K}$.

First, since $\boldsymbol{\mathcal{R}}$ is diagonal, the Sachs equations \eqref{eq:Sachs} only consist of the following two decoupled ordinary differential equations
\begin{align}
\label{eq:Sachs_Kottler_1}
\frac{\dd^2 \xi_1}{\dd v^2} 
&= - \mathcal{R}[r(v)] \, \xi_1(v) , \\
\label{eq:Sachs_Kottler_2}
\frac{\dd^2 \xi_2}{\dd v^2} 
&= + \mathcal{R}[r(v)] \, \xi_2(v) .
\end{align}
Clearly, the decoupling implies that the off-diagonal terms of $\boldsymbol{\mathcal{C}}\e{K}$ and $\boldsymbol{\mathcal{D}}\e{K}$ vanish,
\begin{equation}
{\cal C}\h{K}_{12} = {\cal C}\h{K}_{21}
= {\cal D}\h{K}_{12} = {\cal D}\h{K}_{21} = 0.
\end{equation}

The calculation of the diagonal coefficients requires us to integrate Eqs. \eqref{eq:Sachs_Kottler_1} and \eqref{eq:Sachs_Kottler_2}. This cannot be performed analytically because there is no exact expression for $r$ as a function of $v$ along the null geodesic. Indeed, we can write $v$ as a function of $r$ from Eq.~(\ref{e.udev}) but this relation is not invertible by hand.

Nevertheless, we are able to perform the integration perturbatively in the regime where $\eps_2/\eps_1 \ll \eps_1 \ll 1$, the relevance of which shall be justified by the orders of magnitude discussed in the next section. Solving Eq.~(\ref{e.udev}) at leading order in $\eps_1$, $\eps_2$ leads to
\begin{equation}
u(v) = 
\frac{\eps_1}{\sqrt{1 + (v-v\e{m})^2/\Delta v^2}}
+ \mathcal{O} \left( \eps_1^2, \frac{\eps_2}{\eps_1} \right)
\label{eq:approximate_u_v}
\end{equation}
with $\Delta v \equiv b/E$, and where $v\e{m}$ denotes the value of the affine parameter $v$ at the point of minimal approach. Equation~\eqref{eq:Sachs_Kottler_1} then becomes, at leading order in $\eps_1,\eps_2$, and using the dimensionless variable $w \equiv (v-v\e{m})/\Delta v$,
\begin{equation}
\frac{\dd^2 \xi_1}{\dd w^2} 
= - \frac{3\,\eps_1}{2} 
\, \left( \frac{1}{1 + w^2} \right) ^{5/2} \xi_1 .
\label{eq:approx_diff_xi_1}
\end{equation}
The perturbative resolution of Eq. \eqref{eq:approx_diff_xi_1} from $v\e{init}$ to $v$ finally leads to
\begin{multline}
\mathcal{C}\h{K}_{11}
= 1 - \frac{3\,\eps_1}{2} 
\Big[ -B'(w\e{init}) (w - w\e{init}) \\
+ B(w) - B(w\e{init}) \Big] + \mathcal{O}\left( \eps_1^2, \frac{\eps_2}{\eps_1} \right),
\label{eq:C11_Kottler}
\end{multline}
and
\begin{multline}
\mathcal{D}\h{K}_{11}
= (v - v\e{init})
+ \frac{3\,\eps_1}{2} \, \Delta v \,
\Big\{
	w\e{init}
	\big[ 
	B(w)-B(w\e{init})\\
	-B'(w\e{init}) (w-w\e{init})
	\big]
	- \big[ 
		C(w) - C(w\e{init})\\
		- C'(w\e{init})(w-w\e{init})
	\big] 
\Big\} + \mathcal{O}\left( \eps_1^2, \frac{\eps_2}{\eps_1} \right),
\label{eq:D11_Kottler}
\end{multline}
where the functions $B$ and $C$ are given by
\begin{equation}
B(w) \equiv \frac{1+2\,w^2}{3\sqrt{1+w^2}}
\quad \text{and} \quad
C(w) \equiv \frac{-w}{ 3\sqrt{1+w^2} } .
\end{equation}
The expressions of $\mathcal{C}\h{K}_{22}$ and $\mathcal{D}\h{K}_{22}$ are respectively obtained from Eqs.~\eqref{eq:C11_Kottler} and \eqref{eq:D11_Kottler} by turning $\eps_1$ into $-\eps_1$.

Note that in the limit $\eps_1,\eps_2/\eps_1 \rightarrow 0$, i.e. $b \rightarrow \infty$ and $\Lambda=0$, we find $\boldsymbol{\mathcal{C}}=\boldsymbol{1}$ and $\boldsymbol{\mathcal{D}}=(v-v\e{init})\boldsymbol{1}$, which are the expected expressions in Minkowski spacetime.

\subsection{Practical implementation}
\label{subsec:practical_implementation}

This section has described the complete resolution of the equations for light propagation in a Swiss-cheese universe. All the results are included in a Mathematica program \texttt{OneHole} which takes, as input, the observation conditions and the properties of the hole; and returns $\mathcal{E}\e{in}$, $k\e{in}$ and $\boldsymbol{\mathcal{W}}(v\e{source};v\e{obs})=\boldsymbol{\mathcal{W}}\e{K}(v\e{in};v\e{out})\cdot\boldsymbol{\mathcal{W}}\e{FL}(v\e{out};v\e{obs})$. For simplicity, this program has been written assuming that \emph{the FL region has Euclidean spatial sections} ($K=0$).

Iterating \texttt{OneHole} allows us to propagate a light signal back to an arbitrary emission event. Eventually, the redshift~$z$ is obtained by comparing the wave vector at emission and reception; and the luminosity distance is extracted from the block~$\boldsymbol{\mathcal{D}}(v\e{source} ; v\e{obs})$ of the Wronski matrix~$\boldsymbol{\mathcal{W}}(v\e{source};v\e{obs})$, according to
\begin{equation}
D\e{L}=(1+z)^2 \sqrt{\det \boldsymbol{\mathcal{D}}(v\e{source};v\e{obs})} .
\end{equation}

Note finally that, when iterating \texttt{OneHole}, we must also rotate the Sachs basis~$(s_1,s_2)$, to take into account that the plane of motion differs for two successive holes.

\section{Effect of one hole}
\label{sec:results_one_hole}

Our method is first applied to a Swiss cheese with a single hole. The purpose is to study the effects on the redshift and luminosity distance---for the light emitted by a standard candle---due to the presence of the hole.

\subsection{Numerical values and ``opacity'' assumption}
\label{subsec:numerical_values}

The mass $M$ of the clump inside the hole depends on what object it is supposed to model. The choice must be driven by the typical scales probed by the light beams involved in supernova observations. As discussed in the introduction the typical width of such beams is $\sim \text{AU}$; for comparison the typical interstellar distance within a galaxy is $\sim \text{pc}$. Hence, SN beams are sensitive to the very fine structure of the Universe, including the internal content of galaxies. This suggests that the clump inside the hole should represent a star, so that the natural choice should be $M \sim M_\odot$. Unfortunately, we cannot afford to deal with such a fine description, for numerical reasons.

Instead, the clump is chosen to stand for a gravitationally bound system, such as a galaxy ($M \sim 10^{11} M_\odot$), or a cluster of galaxies ($M \sim 10^{15} M_\odot$). By virtue of Eq.~\eqref{e.massK}, the corresponding hole radii are respectively $r\e{h} \sim 1~\text{Mpc}$ and $r\e{h} \sim 20\,\text{Mpc}$. It is important to note that this choice keeps entirely relevant as far as the light beam does not enter the clump (so that its internal structure does not matter), that is, as long as
\begin{equation}
b > b\e{min} \approx r\e{phys} ,
\label{eq:cutoff_b}
\end{equation}
where $r\e{phys}$ is the physical size of the clump. For a galaxy $r\e{phys} \sim 10\,\text{kpc}$, and for a cluster $r\e{phys} \sim 1\,\text{Mpc}$. We choose to work under the assumption of Eq.\eqref{eq:cutoff_b}, in other words \emph{we proceed as if the clumps were opaque spheres}.

In the case of galactic clumps this ``opacity'' assumption can be justified by the three following arguments (in the case of clusters, however, it is highly questionable).
\begin{description}
\item[Statistics.] Since $r\e{phys} \ll r\e{h}$ the cross section of the clumps is very smal; thus we expect that \emph{most} of the observations satisfy the condition \eqref{eq:cutoff_b}.
\item[Screening.] A galaxy standing on the line of sight can simply be bright enough to flood a SN located behind it. For comparison, the absolute magnitude of a galaxy ranges from $-16$ to $-24$ \cite{GalaxiesCosmologie}, while for a SN it is typically $-19.3$ \cite{SCP2011}.
\item[Strong lensing.] A light beam crossing a galaxy enters the strong lensing regime, because the associated Einstein radius is $r\e{E}\sim\sqrt{r\e{S}D\e{A,SN}} \lesssim 10\,\text{kpc}\sim r\e{phys}$. In this case, we expect a significant magnification of the SN which could be isolated, or even removed during data processing.
\end{description}
The ``opacity'' assumption is at the same time a key ingredient and  a limitation of our approach.

\begin{figure}[h!]
\centering
\includegraphics[width=\columnwidth]{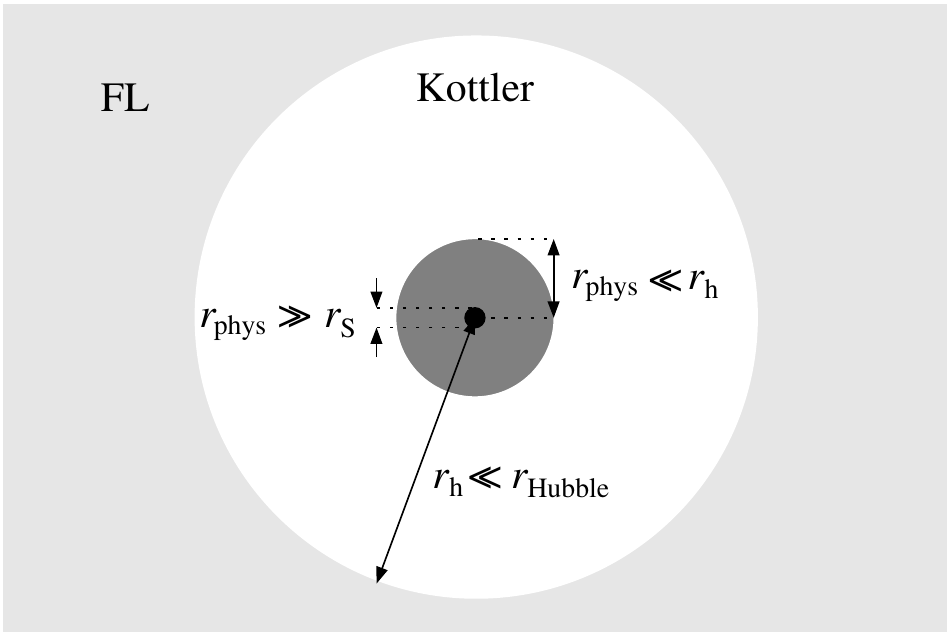}
\caption{Geometry and hierarchy of distances for a typical Swiss-cheese hole: $r\e{S} \ll r\e{phys} \ll r\e{h} \ll r\e{Hubble}$.}
\label{fig:hierarchy_hole}
\end{figure}

The various distance scales involved in the model are clearly separated. The resulting hierarchy is depicted in Fig.~\ref{fig:hierarchy_hole}, and the typical orders of magnitude are summarized in Table~\ref{tab:oom}. The latter includes the small parameters $\eps_1=r\e{S}/b$ and $\eps_2=\Lambda r\e{S}^2/3\sim (r\e{S}/r\e{Hubble})^2$. Their values justify \textit{a posteriori} the perturbative expansion performed in \S~\ref{paragraph:perturbative_Wronski}, where we assumed that $\eps_2/\eps_1 \ll \eps_{1} \ll 1$. In fact, one can show from Eq.~\eqref{e.massK} that $\eps_2 \sim \eps_{1,\text{min}}^3$.

\begin{table}[h!]
\centering
\begin{tabular}{|c||ccccc|}
\hline 
Type & $r\e{S}$ (pc) & $r\e{phys}$ (kpc) & $r\e{h}$ (Mpc) & $\eps_1$ & $\eps_2$ \\ 
\hline 
Galaxy & $10^{-2}$ & 10 & 1 & $10^{-8}$--$10^{-6}$ & $10^{-23}$ \\ 
Cluster & $100$ & 1000 & 20 & $10^{-6}$--$10^{-4}$ & $10^{-15}$ \\ 
\hline 
\end{tabular}
\caption{Typical orders of magnitude for galaxylike ($M\sim 10^{11}M_\odot$) and clusterlike ($M\sim 10^{15}M_\odot$) Swiss-cheese holes.}
\label{tab:oom}
\end{table}

In this section and the next one, we \emph{temporarily} set for simplicity the cosmological constant to zero. The FL region is therefore characterized by the Einstein--de Sitter (EdS) cosmological parameters
\begin{equation}
\Omega_{\rm m} = 1, \qquad
\Omega_{K} = 0, \qquad
\Omega_{\Lambda} = 0.
\end{equation}
The effect of the cosmological constant will be studied in detail in Sec.~\ref{sec:cosmological_consequences}. The value of the Hubble parameter is fixed to $H_0=h \times 100\,\text{km/s/Mpc}$, with $h=0.72$.

\subsection{Setup}
\label{subsec:setup}

\begin{figure}[!h]
\centering
\includegraphics[width=\columnwidth]{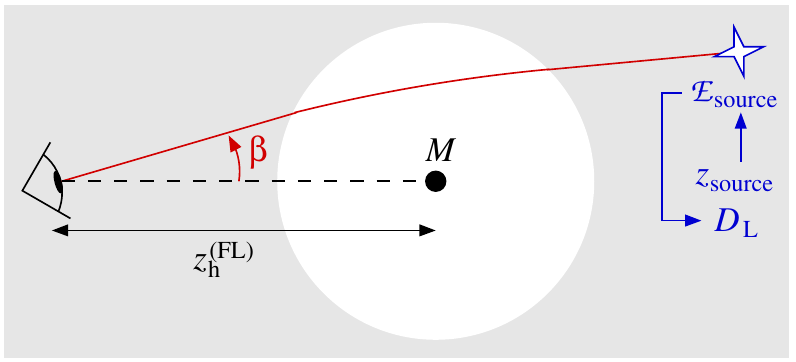}\\
\vspace*{0.5cm}
\includegraphics[width=\columnwidth]{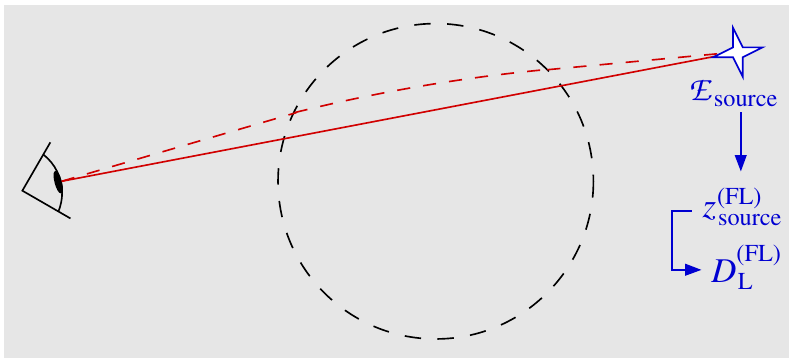}
\caption{Setup for evaluating the effect of one hole on the redshift and luminosity distance.}
\label{fig:effect_one_hole}
\end{figure}

In order to study the corrections to the redshift $z$ and luminosity distance $D\e{L}$, due to the presence of the hole, we consider the situation depicted in Fig.~\ref{fig:effect_one_hole}.

Our method is the following. We first choose the mass~$M$ inside the hole and the redshift~$z\e{source}$ of the source. We then fix the comoving distance between the observer and the center of the hole, in terms of the cosmological (FL) redshift~$z\e{h}\h{(FL)}$ of the latter. To finish, we choose a direction of observation, defined by the angle~$\beta$ between the line of sight and the line connecting the observer to the center of the hole.

Given those parameters, the light beam is propagated (in presence of the hole) until the redshift reaches $z\e{source}$. We obtain the emission event $\mathcal{E}\e{source}$ and the luminosity distance $D\e{L}$. We then compute $z\e{source}\h{(FL)}$ and $D\e{L}\h{(FL)}$ by considering a light beam that propagates from $\mathcal{E}\e{source}$ to the observer without the hole (bottom panel of Fig.~\ref{fig:effect_one_hole}).

\subsection{Corrections to the redshift}

\subsubsection{Numerical results}

\begin{figure}[!h]
\centering
\includegraphics[width=\columnwidth]{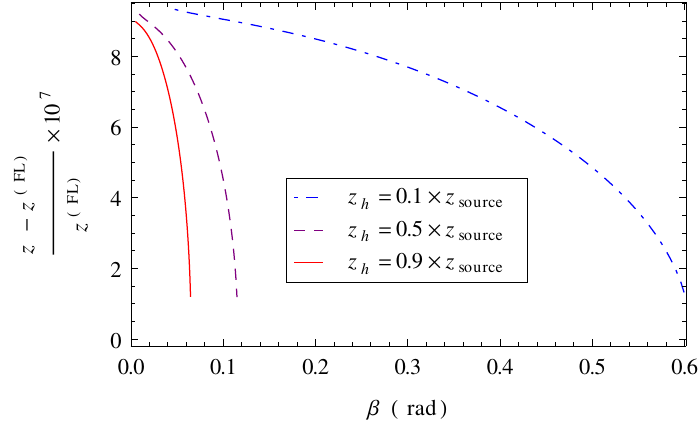}\\
\includegraphics[width=\columnwidth]{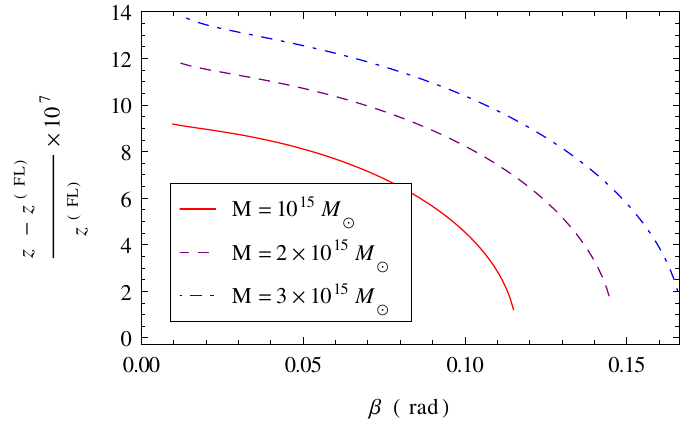}
\caption{Relative correction to the redshift $z$, due to the hole in the line of sight, as a function of the direction of observation~$\beta$, for a source at $z\e{source}=0.05$. Top panel: The mass of the hole is $M=10^{15} M_\odot$, and three positions between the source and the observer are tested, $z\e{h}\h{(FL)}/z\e{source}=0.1$ (blue, dot-dashed), $0.5$ (purple, dashed), and $0.9$ (red, solid). Bottom panel: The hole is at $z\e{h}\h{(FL)}=0.5\,z\e{source}$ and three values for the mass are tested, $M/10^{15}M_\odot=3$ (blue, dot-dashed), $2$ (purple, dashed), and $1$ (red, solid).}
\label{fig:onehole_z_beta}
\end{figure}

The effect of the hole on the redshift is quantified by
\begin{equation}
\delta z \equiv \frac{z-z\h{(FL)}}{z\h{(FL)}} ,
\end{equation}
where we used the short notation $z$ instead of $z\e{source}$. Figure~\ref{fig:onehole_z_beta} shows the evolution of $\delta z$ with $\beta$, for $z\e{source}=0.05$ and various hole positions and masses. We have chosen $M \sim 10^{15} M_\odot$ because the effect is more significant and displays fewer numerical artifacts than for $M\sim 10^{11} M_\odot$.

\begin{figure}[!h]
\centering
\includegraphics[width=0.97\columnwidth]{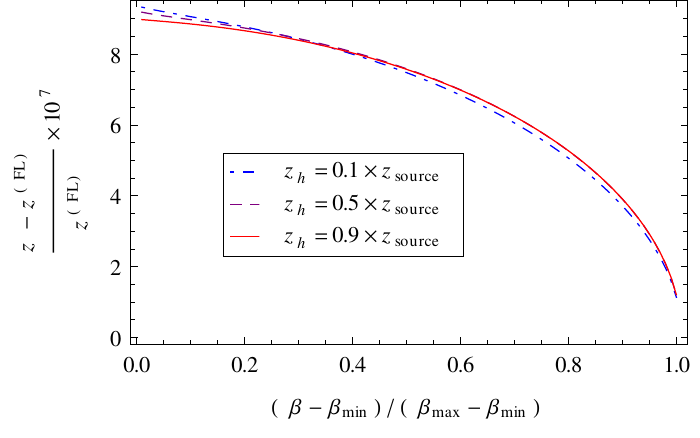}\\
\includegraphics[width=\columnwidth]{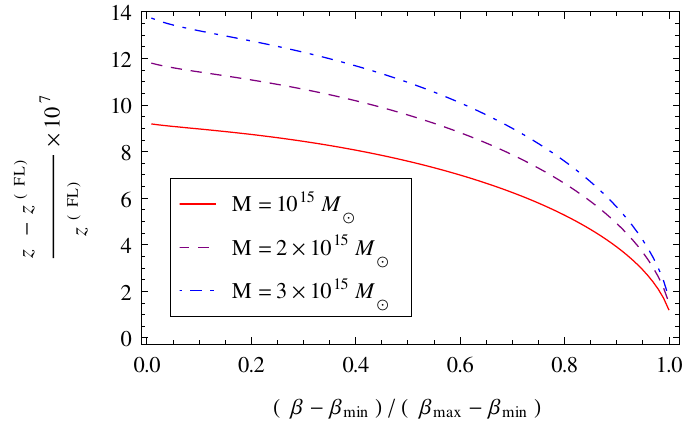}
\caption{Same as Fig. \ref{fig:onehole_z_beta}, but plotted in terms of the centered and normalized observation direction $(\beta-\beta\e{min})/(\beta\e{max}-\beta\e{min})$.}
\label{fig:onehole_z_deltabeta}
\end{figure}

We only consider directions of observation such that the light beam crosses the hole. Thus, $\beta\e{min}<\beta<\beta\e{max}$ where $\beta\e{min}$ and $\beta\e{max}$ depend on the physical cutoff~$r\e{phys}$, the radius~$r\e{h}$ of the hole, and its distance to the observer~$z\e{h}\h{(FL)}$. Those dependences can be eliminated by plotting $\delta z$ as a function of $(\beta-\beta\e{min})/(\beta\e{max}-\beta\e{min})$ instead of $\beta$, as displayed in Fig. \ref{fig:onehole_z_deltabeta}.

As expected, $\delta z$ tends to zero when $\beta$ approaches $\beta\e{max}$ (light ray tangent to the hole boundary). We notice that $\delta z$ does not significantly depend on the distance between the observer and the hole. However, the effect clearly grows with the mass of the hole.

\subsubsection{Analytical estimation of the effect}

The correction in redshift due to hole can be understood as an integrated Sachs-Wolfe effect (see e.g. Chapter~7 of Ref.~\cite{pubook}). As the boundary of the hole grows with time (see Fig. \ref{fig:deflection}), the light signal undergoes a stronger gravitational potential at entrance than at exit. That induces a gravitational redshift $\delta z\e{grav}$ which adds to the cosmological one, and reads
\begin{equation}
1 + \delta z\e{grav} = \frac{k^t\e{in}}{k^t\e{out}} = \frac{A(r\e{out})}{A(r\e{in})} .
\label{eq:def_deltazgrav}
\end{equation}

The order of magnitude of $\delta z\e{grav}$ can be evaluated as follows. Let $\delta r = r\e{out}-r\e{in}$ be the increase of the radius of the hole between entrance and exit. The expansion dynamics implies $\delta r \sim \sqrt{\eps_1} \Delta t$, where $\Delta t = t\e{out} - t\e{in} \sim r\e{in}$, $r\e{out}$ is the time spent by the photon inside the hole. Using Eq. \eqref{eq:def_deltazgrav}, we conclude that
\begin{equation}
\delta z\e{grav} \sim \eps_1^{3/2} .
\end{equation}
For $M=10^{15} M_\odot$ (clusterlike hole), the numerical values given in Table~\ref{tab:oom} yield $\delta z\e{grav,max} \sim 10^{-6}$. This order of magnitude is compatible with the full numerical integration displayed in Figs.~\ref{fig:onehole_z_beta} and \ref{fig:onehole_z_deltabeta}.

Such an analytical estimate enables us to understand why $\delta z$ increases with $M$, that is, with the size of the hole. Indeed, the bigger the hole, the longer the photon travel time so that the hole has more time to grow, and finally $A(r\e{out}) - A(r\e{in})$ is larger.

\subsection{Corrections to the luminosity distance}

\begin{figure}[!h]
\centering
\includegraphics[width=\columnwidth]{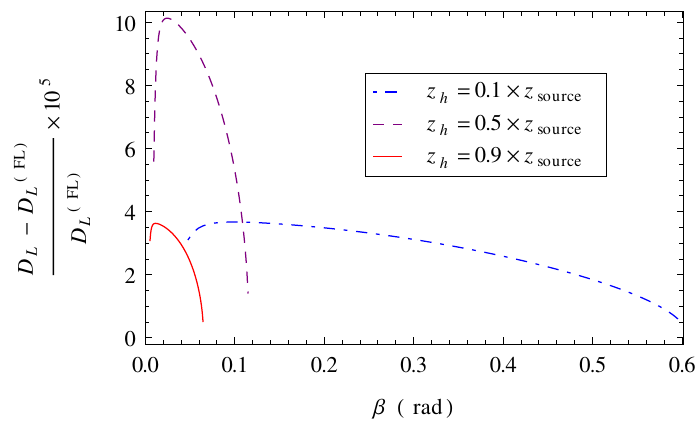}\\
\includegraphics[width=\columnwidth]{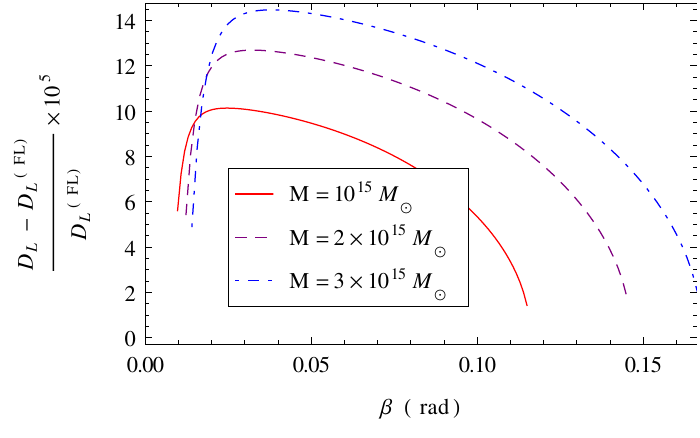}
\caption{Relative correction to the luminosity distance $D\e{L}$, due to the hole in the line of sight, as a function of the direction of observation~$\beta$, for a source at $z\e{source}=0.05$. Top panel: The mass of the hole is $M=10^{15} M_\odot$, and three positions between the source and the observer are tested, $z\e{h}\h{(FL)}/z\e{source}=0.1$ (blue, dot-dashed), $0.5$ (purple, dashed), and $0.9$ (red, solid). Bottom panel: The hole is at $z\e{h}\h{(FL)}=0.5\,z\e{source}$ and three values for the mass are tested, $M/10^{15}M_\odot=3$ (blue, dot-dashed), $2$ (purple, dashed), and $1$ (red, solid).}
\label{fig:onehole_DL_beta}
\end{figure}

\begin{figure}[!h]
\centering
\includegraphics[width=\columnwidth]{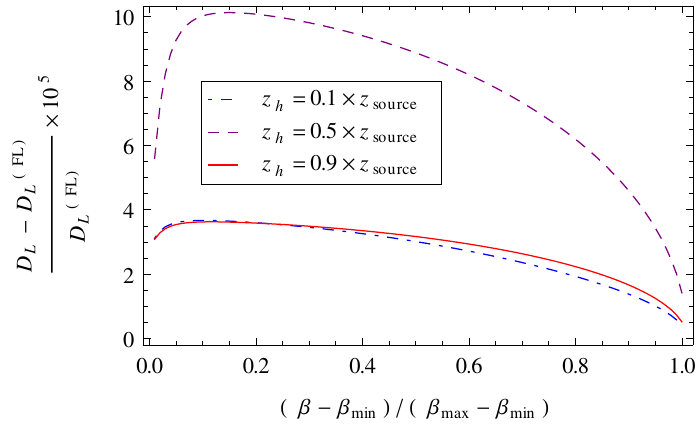}\\
\includegraphics[width=\columnwidth]{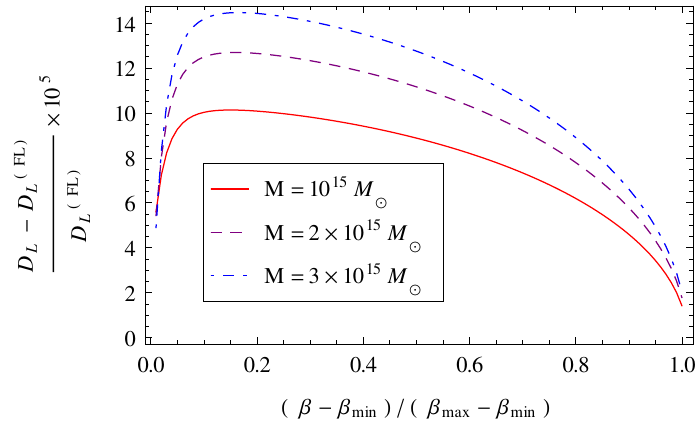}
\caption{Same as Fig. \ref{fig:onehole_DL_beta}, but plotted in terms of the centered and normalized observation direction $(\beta-\beta\e{min})/(\beta\e{max}-\beta\e{min})$.}
\label{fig:onehole_DL_deltabeta}
\end{figure}

The effect of the hole on the luminosity distance can be characterized in a similar way by
\begin{equation}
\delta D\e{L} \equiv \frac{D\e{L}-D\e{L}\h{(FL)}}{D\e{L}\h{(FL)}} .
\end{equation}
The associated results, in the same conditions as in the previous paragraph, are displayed in Figs.~\ref{fig:onehole_DL_beta} and \ref{fig:onehole_DL_deltabeta}.

We notice that $\delta D\e{L}$ is maximum if the hole lies halfway between the source and the observer, which is indeed expected since the lensing effects scale as 
\begin{equation}
\frac{D\e{A}(\text{observer, lens})
		\times 
		D\e{A}(\text{lens, source})}
	{D\e{A}(\text{observer, source})},
\end{equation}
which typically peaks for $z_{\rm lens} \approx z_{\rm source}/2$. The maximal amplitude of the correction is of order $10^{-4}$, for masses ranging from $10^{15}M_\odot$ to $3\times10^{15}M_\odot$. Just as for the redshift, the effect increases with the size of the hole.

Note that $\delta D\e{L}$ can be related to the relative magnification~$\mu$, frequently used in the weak-lensing formalism, and defined by
\begin{equation}
\mu \equiv \left( \frac{D\e{A}\h{(FL)}}{D\e{A}}  \right)^2
= \left( \frac{1+z}{1+z\h{(FL)}} \right)^4 
\left( \frac{D\e{L}\h{(FL)}}{D\e{L}} \right)^2.
\end{equation}
Hence, if the correction on $z$ is negligible compared to the one of $D\e{A}$, then the relation between $\delta D\e{L}$ and $\mu$ is
\begin{equation}
\delta D\e{L} \approx \frac{1}{\sqrt{\mu}} - 1 .
\end{equation}

\subsection{Summary}

The presence of a single hole between the source and the observer induces both a correction in redshift and luminosity distance. For a hole with mass $M \sim 10^{15} M_\odot$, the relative amplitudes of those corrections are $\delta z \sim 10^{-7}$--$10^{-6}$ and $\delta D\e{L} \sim 100 \, \delta z$. The same study for $M \sim 10^{11} M_\odot$ leads to similar results with $\delta z \sim 10^{-10}$--$10^{-9}$. Therefore, the effects of a single hole seem negligible.

\section{Effect of several holes}
\label{sec:results_several_holes}

We now investigate a Swiss-cheese model containing many holes arranged on a regular lattice. Again, in this entire section, the cosmological parameters characterizing the FL region are the EdS ones.

\subsection{Description of the arrangement of holes}

\subsubsection{Smoothness parameter}
\label{subsubsec:smoothnes_parameter}

The smoothness of the distribution of matter within a Swiss cheese can be quantified by a parameter $f$ constructed as follows. Choose a region of space with---comoving or physical---volume~$V$, where $V^{1/3}$ is large compared to the typical distance between two holes. Thus, this volume contains many holes, the total volume of which is $V\e{holes}$, while the region left with homogeneous matter occupies a volume $V_{\rm FL} = V-V\e{holes}$. We define the smoothness parameter by
\begin{equation}
f \equiv \underset{V\rightarrow\infty}{\rm lim} 
	\frac{V\e{FL}}{V}.
\end{equation}
In particular, $f=1$ corresponds to a Swiss cheese with no hole---that is, perfectly smooth---while $f=0$ corresponds to the case where matter is under the form of clumps. Of course, $f$ also characterizes the ratio between the energy density of the continuous matter and the mean energy density.

\subsubsection{Lattice}

\begin{figure}[!h]
\centering
\includegraphics[
clip=true,trim=0.1cm 0.1cm 30cm 0.1cm,
width=0.49\columnwidth]{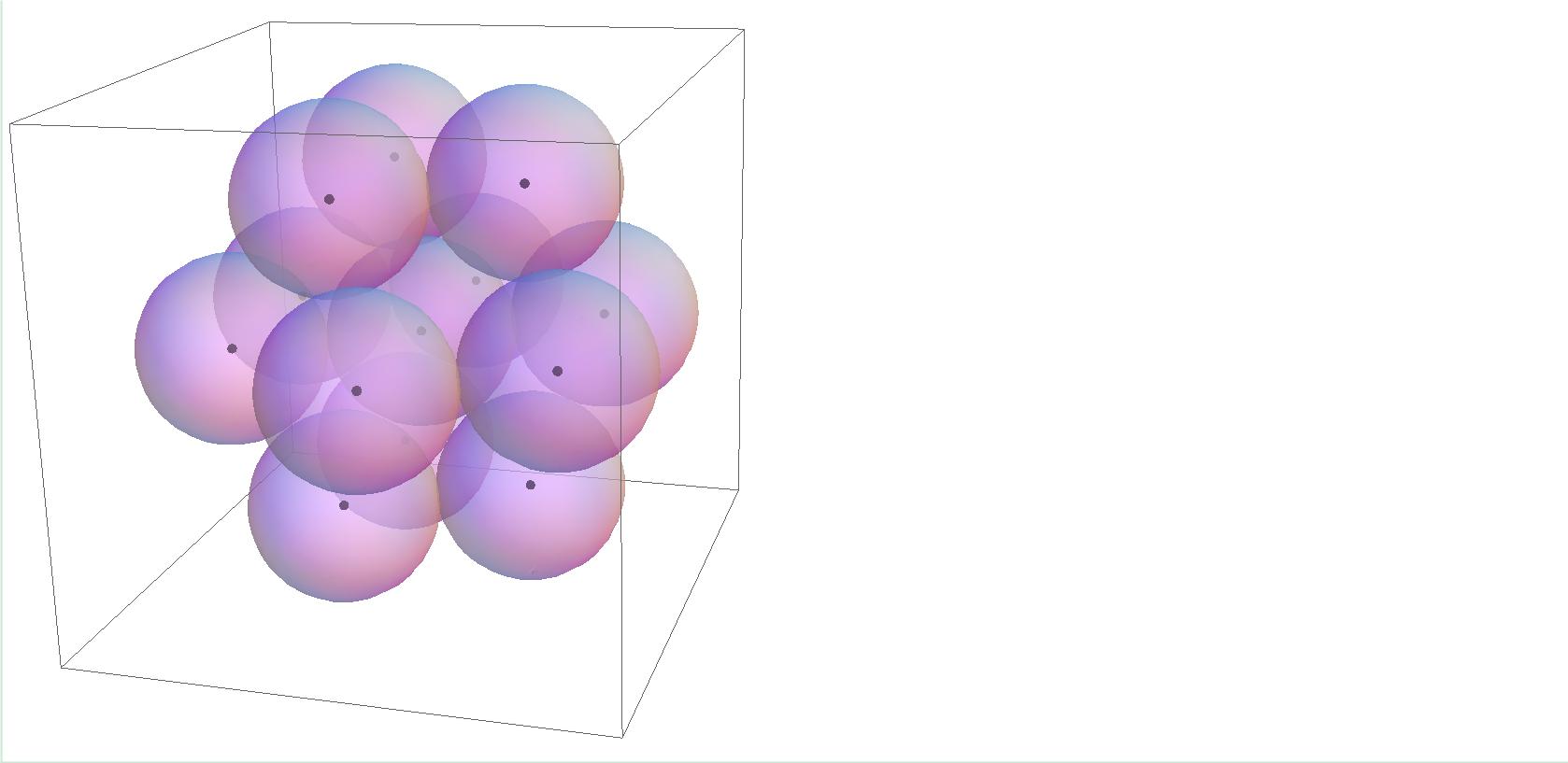}
\includegraphics[
clip=true,trim=0.1cm 0.1cm 30cm 0.1cm,
width=0.49\columnwidth]{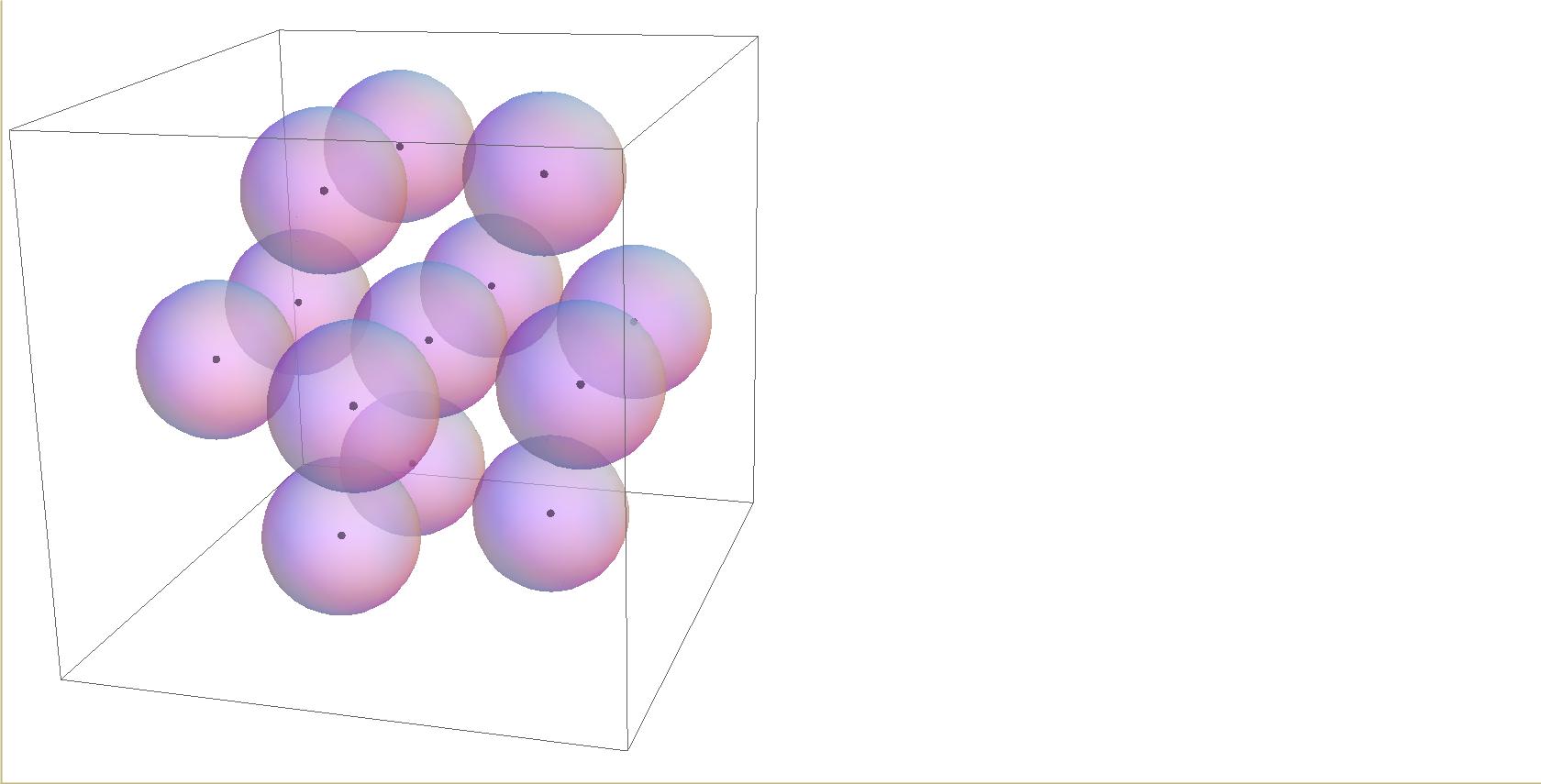}
\caption{Hexagonal lattice of identical holes. On the left, the arrangement is close-packed, so that the smoothness parameter is $f=f\e{min}\approx 0.26$. On the right, $f=0.7$.}
\label{fig:hexagonal_lattice}
\end{figure}

We want to construct a Swiss cheese for which the smoothness parameter is as small as possible. If all holes are identical, this close-packing problem can be solved by using, for instance, a hexagonal lattice. The corresponding arrangement is pictured in Fig. \ref{fig:hexagonal_lattice}. The minimal value of the smoothness parameter is in this case
\begin{equation}
f\e{min} = 1 - \frac{\pi}{3\sqrt{2}} \approx 0.26 .
\end{equation}
In order to reach a smoothness parameter smaller than $f\e{min}$, one would have to insert a second family of smaller holes. By iterating the process, one can in principle make $f$ as close as one wants to zero.

\subsection{Observations in a unique line of sight}
\label{subsec:single_obs_direction}

We now focus on the corrections to the redshift and luminosity distance of a source whose light travels through the Swiss-cheese universe described previously. We study the influence of (a) the distance between the source and the observer, (b) the smoothness parameter $f$, and (c) the mass $M$ of the holes.

\subsubsection{Setup}

After having chosen the parameters $(f,M)$ of the model, we arbitrarily choose the spatial position of the observer in the FL region, and fix its direction of observation. The method is then identical to the one of Sec.~\ref{sec:results_one_hole}. The light beam is propagated from the observer until the redshift reaches the one of the source, $z$. The ending point defines the emission event $\mathcal{E}\e{source}$. We emphasize that only emission events occurring in the FL region are considered in this article.

\subsubsection{Influence of the smoothness parameter}

\begin{figure}[!h]
\centering
\includegraphics[width=\columnwidth]{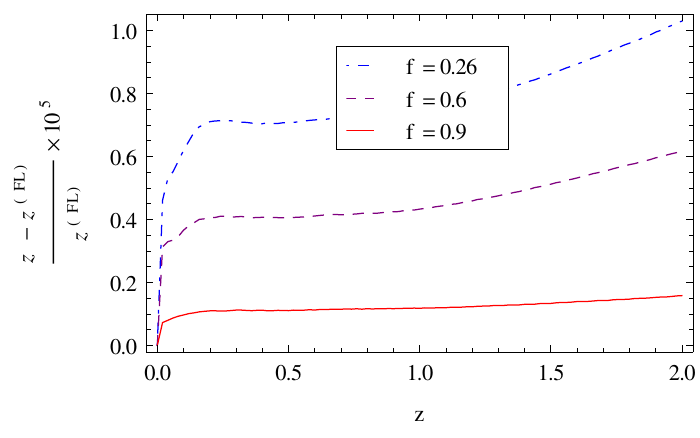}\\
\includegraphics[width=\columnwidth]{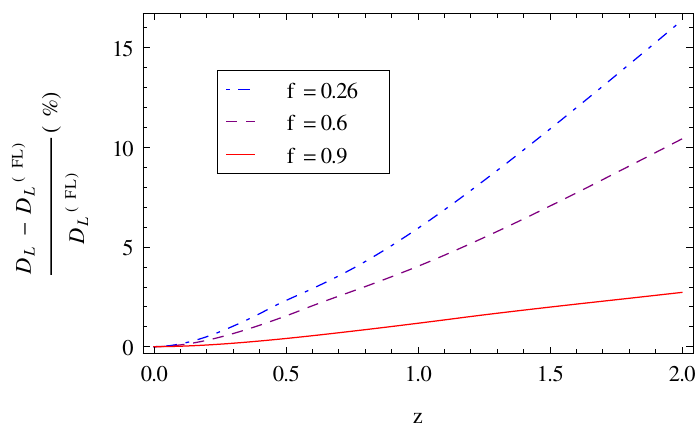}
\caption{Relative corrections to the redshift~$z$ (top panel) and luminosity distance~$D\e{L}$ (bottom panel) as functions of $z$, for an arbitrary light beam traveling through a Swiss-cheese universe. All holes are identical, their mass is $M=10^{11} M_\odot$. Three different smoothness parameters are tested: $f=0.26$ (blue, dot-dashed), $0.6$ (purple, dashed), and $0.9$ (red, solid).}
\label{fig:manyhole_f_varies}
\end{figure}

In this paragraph, the mass of every hole is fixed to $M=10^{11} M_\odot$ (galactic holes). The relative corrections to the redshift $\delta z$ and luminosity distance $\delta D\e{L}$, as functions of the redshift $z$ of the source, have been computed and are displayed in Fig. \ref{fig:manyhole_f_varies} for different values of the smoothness parameter $f$.

While the corrections to the redshift remain small---typically $\delta z < 10^{-5}$---the cumulative effect of lensing on the luminosity distance is significant. For instance, a source at $z \sim 1.5$ would appear $10\,\%$ farther in a Swiss cheese with $f=0.26$, than in a strictly homogeneous universe. Both $\delta z$ and $\delta D\e{L}$ increase with $z$ and decrease with $f$, as intuitively expected. Thus, the more holes, the stronger the effect. As examples, the light beam crosses $\sim 300$ holes for $(f=0.26,z=0.1)$ or $(f=0.9,z=1)$, but it crosses $\sim 2000$ holes for $(f=f\e{min},z=1)$.

\subsubsection{Influence of the mass of the holes}

\begin{figure}[!h]
\centering
\includegraphics[width=\columnwidth]{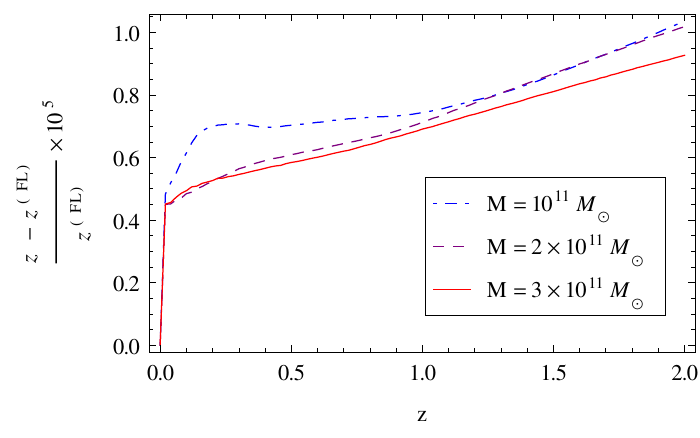}\\
\includegraphics[width=\columnwidth]{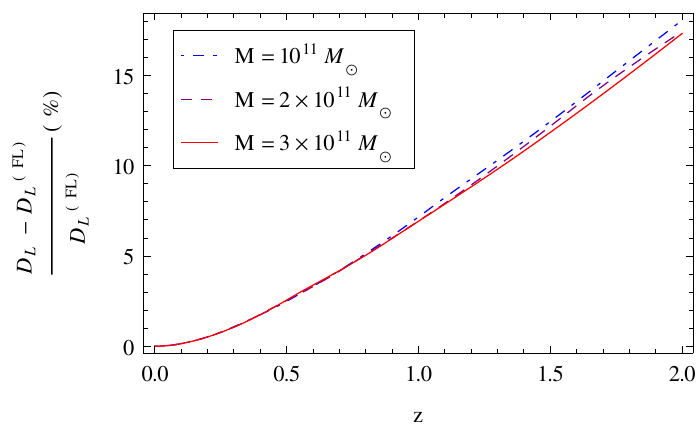}
\caption{Same as Fig. \ref{fig:manyhole_f_varies} but with $f=0.26$ and three different values for the masses: $M/10^{11} M_\odot = 1$ (blue, dot-dashed), $2$ (purple, dashed), and $3$ (red, solid).}
\label{fig:manyhole_M_varies}
\end{figure}

We now set the smoothness parameter to its minimal value $f\e{min} \approx 0.26$, and repeat the previous analysis for various hole masses. The results are displayed in Fig.~\ref{fig:manyhole_M_varies}.

We conclude that neither $\delta z$ nor $\delta D\e{L}$ depends significantly on $M$, that is, on the size of the holes. Thus, what actually matters is not the number of holes intersected by the beam, but rather the total time spent inside holes.

\subsection{Statistical study for random directions of observation}
\label{subsec:statistical_study}

The previous study was restricted to a single line of sight, but since a Swiss-cheese universe is not strictly homogeneous, the corrections to $z$ and $D\e{L}$ are expected to vary from one line of sight to another. As pointed out by e.g. Refs.~\cite{Vanderveld:2008vi,Szybka:2010ky}, such a restrictive analysis can lead to overestimate the mean corrections induced by inhomogeneities. Besides, as stressed by Ref.~\cite{cemuu}, the dispersion of the data is crucial for interpreting SN observations. Hence, the conclusions of the previous subsection need to be completed by a statistical study, with randomized directions of observation.

Since the effect on the redshift is observationally negligible, we focus on the luminosity distance. After having set the parameters $(f,M)$ of the model, we fix the position of the observer in the FL region. Then, for a given redshift $z$, we consider a statistical sample of $N\e{obs}$ randomly distributed directions of observation~$\vec{d}\in\mathcal{S}^2$, and compute $\delta D\e{L}(z,\vec{d})$ for each one.

\begin{figure}[!h]
\centering
\includegraphics[width=\columnwidth]{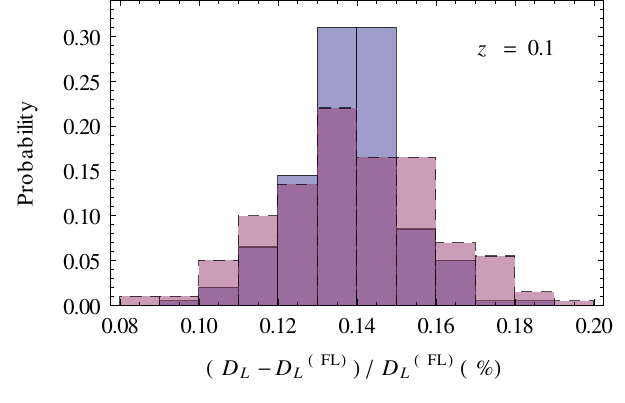}\\
\includegraphics[width=\columnwidth]{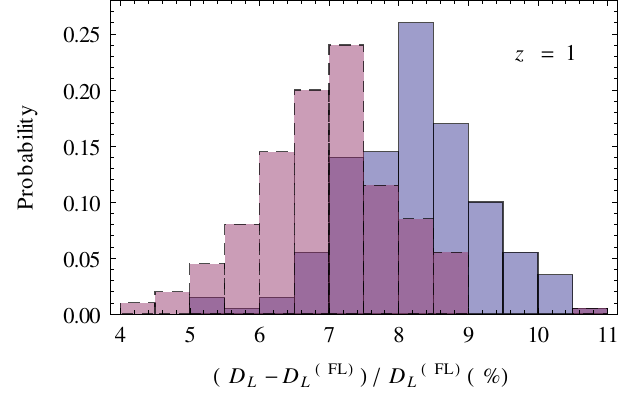}
\caption{Probability distribution of the relative correction to the luminosity distance for random directions of observation, at $z=0.1$ (top panel) and $z=1$ (bottom panel). The smoothness parameter is $f=f\e{min}$ and two different values for the masses of the holes are tested: $M=10^{11}M_\odot$ (blue, solid) and $M=10^{15} M_\odot$ (purple, dashed).}
\label{fig:histograms}
\end{figure}

Figure~\ref{fig:histograms} shows the probability distribution of $\delta D\e{L}$ for sources at redshifts $z=0.1$ (top panel) and $z=1$ (bottom panel). We compare two Swiss-cheese models with the same smoothness parameter $f=f\e{min}$ but with different values for the masses of their holes ($M=10^{11} M_\odot$ and $10^{15} M_\odot$). The histograms of Fig.~\ref{fig:histograms} are generated from statistical samples which contain $N\e{obs}=200$ directions of observation each.

\begin{figure}[!h]
\centering
\includegraphics[width=\columnwidth]{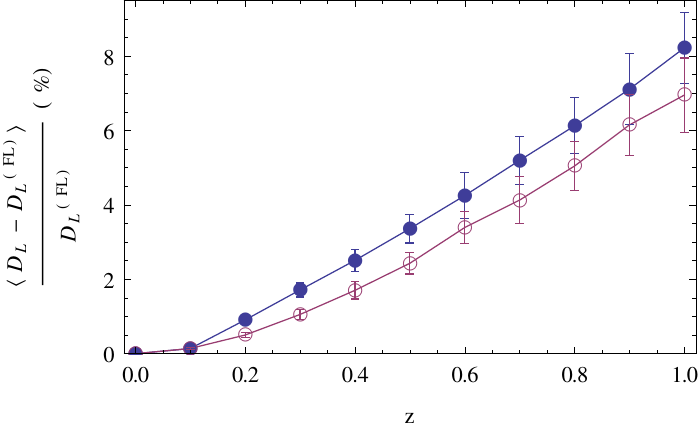}
\caption{Evolution, with redshift $z$, of the relative correction to the luminosity distance averaged over $N\e{obs}=200$ random directions of observation. Error bars indicate the dispersion $\sigma_{\delta D\e{L}}$ around the mean correction $\langle \delta D\e{L}\rangle$. As in Fig.~\ref{fig:histograms}, we compare Swiss-cheese models with $M=10^{11}M_\odot$ (blue, filled markers) and $M=10^{15} M_\odot$ (purple, empty markers).}
\label{fig:mean_and_dispersion}
\end{figure}

From the statistical samples, we can compute the mean correction $\langle \delta D\e{L} \rangle(z)$ and its standard deviation $\sigma_{\delta D\e{L}}(z)$, whose evolutions are plotted in Fig. \ref{fig:mean_and_dispersion}.

The results displayed in Fig. \ref{fig:mean_and_dispersion} confirm the conclusions of Subsec.~\ref{subsec:single_obs_direction}. The distance-redshift relation in a Swiss cheese is biased with respect to the one of a purely homogenous universe. This effect is statistically significant, we indeed estimate (empirically) that
\begin{equation}
\langle \delta D\e{L} \rangle (z)
\approx 8 \times \sigma_{\delta D\e{L}}(z).
\end{equation}
The bias slightly decreases with the mass parameter $M$. However, it can be considered quite robust because a variation of 4 orders of magnitude for $M$ only induces a variation of $\sim 10\%$ for the bias.

The intrinsic dispersion of $D\e{L}$, associated with $\sigma_{\delta D\e{L}}$, can be compared with the typical dispersion of the observation. For instance, at $z=1$ the former is~$\sim 1\,\%$, while the latter is estimated to be typically $\sim 10\,\%$ \cite{SNIa}. It follows that the dispersion induced by the inhomogeneity of the distribution of matter remains small compared to the observational dispersion.

\subsection{Summary and discussion}
\label{subsec:summary_discussion}

This section has provided a complete study of the effect of inhomogeneities on the Hubble diagram, investigating both the corrections to the redshift and luminosity distance of standard candles. The Swiss-cheese models are made of identical holes, defined by their mass~$M$, and arranged on a regular hexagonal lattice. The fraction of matter remaining in FL regions defines the smoothness parameter~$f$. For the hexagonal lattice, $f\e{min} \approx 0.26$.

The effect on the redshift is negligible ($\delta z<10^{-5}$), while the correction to the luminosity distance is significant ($\delta D\e{L} > 10\,\%$ at high redshift). Compared to the homogeneous case, sources are systematically demagnified in a Swiss-cheese universe. The effect increases with $z$ and decreases with $f$.

Our results differ from those obtained in Swiss-cheese models with Lema\^itre-Tolman-Bondi (LTB) solutions inside the holes. In the latter case, a source can be either demagnified if light mostly propagates through underdense regions \cite{Brouzakis:2006dj,Marra:2007pm,Clifton:2009nv} (and if the observer is far away from a void, see Ref.~\cite{Anti-lensing}), or magnified otherwise. It has been proven in Refs.~\cite{Brouzakis:2008,Szybka:2010ky,Vanderveld:2011} that the global effect averages to zero when many sources are considered. Hence, LTB holes introduce an additional dispersion to the Hubble diagram, but no statistically significant bias. On the contrary, in the present study, light only propagates through underdense regions, because we only consider light beams which remain far from the hole centers. This assumption has been justified in Subsec.~\ref{subsec:numerical_values} by an ``opacity'' argument. The bias displayed by our results is mostly due to the selection of the light beams which can be considered observationally relevant.

Our results also differ qualitatively from those obtained in the framework of the perturbation theory. In Ref.~\cite{Valageas2000}, the probability density function~$P(\mu)$ of the weak lensing magnification~$\mu$, due to the large scale structure, has been analytically calculated by assuming an initial power spectrum with slope $n=-2$. Just as for LTB Swiss-cheese models, the magnification shows no intrinsic bias (i.e. $\langle \mu \rangle = 1$), but it is shown that $P(\mu)$ peaks at a value $\mu\e{peak}$ slightly smaller than $1$. Hence, a bias of order $\mu\e{peak} - \langle \mu \rangle$, which is typically $1\,\%$ at $z=1$, can emerge from observations because of insufficient statistics. However, this bias is far smaller than the one obtained in our Swiss-cheese model, of order $2\,\delta D\e{L} \sim 15\,\%$ at $z=1$.

Besides, the dispersion around the mean magnification is stronger for perturbation theory ($\sim 10\,\%$) than for both LTB and Kottler Swiss-cheese models ($\sim 2\,\%$).

\section{Cosmological consequences}
\label{sec:cosmological_consequences}

Since the Hubble diagram is modified by the presence of inhomogeneities, the resulting determination of the cosmological parameters must be affected as well.

More precisely, consider a Swiss-cheese universe whose FL regions are characterized by a set of cosmological parameters $(\Omega\e{m},\Omega_{K},\Omega_{\Lambda})$, called \emph{background} parameters in the following. If an astronomer observes SNe in this inhomogeneous universe and constructs the resulting Hubble diagram, but fits it with the usual FL luminosity-redshift relation---that is, assuming that he lives in a strictly homogeneous universe---then he will infer \emph{apparent} cosmological parameters $(\bar{\Omega}\e{m},\bar{\Omega}_{K},\bar{\Omega}_{\Lambda})$ which shall differ from the background ones. Evaluating this difference is the purpose of Subsecs.~\ref{subsec:generating_mock_Hubble_diagrams}, \ref{subsec:determining_cosmological_parameters}, and \ref{subsec:results_cosmo_parameters}.

The natural question which comes after is, assuming that our own Universe is well described by a Swiss-cheese model, what are the \emph{background} cosmological parameters which best reproduce the actual SN observations? This issue is addressed in~Subsec.~\ref{subsec:alternative_fit}.

\subsection{Generating mock Hubble diagrams}
\label{subsec:generating_mock_Hubble_diagrams}

The Hubble diagram observed in a given Swiss-cheese universe is constructed in the following way. We first choose the parameters of the model: $f$, $M$, and the background cosmology $(\Omega\e{m},\Omega_{\Lambda}=1-\Omega\e{m})$.\footnote{Recall that in the practical implementation of the theoretical results (see Subsec.~\ref{subsec:practical_implementation}), we assumed that $K=0$, so that the background cosmology of our Swiss-cheese models is completely determined by $\Omega\e{m}$ or $\Omega_\Lambda$. Nevertheless, the \emph{apparent} curvature parameter $\bar{\Omega}_{K}$ is \textit{a priori} nonzero.} We then fix arbitrarily the position of the observer in the FL region, and we simulate observations by picking randomly the line of sight~$\vec{d}$, the redshift $z \in [0,z\e{max}]$, and we compute the associated luminosity distance $D\e{L}(z,\vec{d})$ as in Sec.~\ref{sec:results_several_holes}. In order to make our mock SNe catalog resemble the SNLS~3 data set \cite{SNLS3}, we choose $z\e{max}=1.4$ and $N\e{obs}=472$.

\begin{figure}[!h]
\centering
\includegraphics[clip=true,trim=0cm 0cm 0.4cm 0cm,width=\columnwidth]{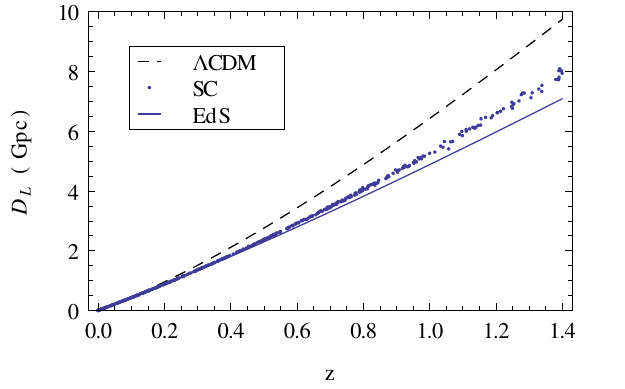}
\caption{Hubble diagram of a Swiss-cheese universe (dots) with $f=f\e{min}$, $M=10^{11} M_\odot$ and EdS background cosmology. For comparison, we also display the distance-redshift relations of purely FL universes, with EdS parameters (blue, solid) and $(\Omega\e{m},\Omega_{K},\Omega_{\Lambda})=(0.3,0,0.7)$---(black, dashed).}
\label{fig:example_Mock_Hubble}
\end{figure}

An example of mock Hubble diagram, corresponding to a Swiss-cheese model with $f=f\e{min}$, $M=10^{11} M_\odot$ and $(\Omega\e{m},\Omega_{K},\Omega_{\Lambda})=(1,0,0)$ is plotted in Fig. \ref{fig:example_Mock_Hubble}. As a comparison, we also displayed $D\e{L}(z)$ for a homogeneous universe with (1) the same cosmological parameters, and (2) with $(\Omega\e{m},\Omega_{K},\Omega_{\Lambda})=(0.3,0,0.7)$.

\subsection{Determining apparent cosmological parameters}
\label{subsec:determining_cosmological_parameters}

The apparent cosmological parameters $\bar{\Omega}\e{m}$, $\bar{\Omega}_{\Lambda}$ and $\bar{\Omega}_{K}=1-\bar{\Omega}\e{m}-\bar{\Omega}_{\Lambda}$ are determined from the mock Hubble diagrams by performing a $\chi^2$ fit. The $\chi^2$ is defined by
\begin{equation}
\chi^2 ( \bar{\Omega}\e{m} , \bar{\Omega}_{\Lambda} )
\equiv
\sum_{i=1}^{472}
\left[
\frac{ \mu_i - \mu\e{FL}
	(z_i \,|\, \bar{\Omega}\e{m},\bar{\Omega}_{\Lambda}) }
{ \Delta \mu_i }
\right]^2 ,
\label{eq:chi2}
\end{equation}
where $\mu$ no longer denotes the magnification, but rather the distance modulus associated with $D\e{L}$, so that
\begin{equation}
\mu \equiv 5\,\log_{10} \left( \frac{D\e{L}}{10\,\text{pc}} \right) .
\end{equation}
In Eq. \eqref{eq:chi2}, $(z_i,\mu_i)$ is the $i$th observation of the simulated catalog. In order to make the analysis more realistic, we have attributed to each data point an observational error bar~$\Delta \mu_i$ estimated by comparison with the SNLS~3 data set~\cite{SNLS3}. Besides, $\mu\e{FL}(z \,|\, \bar{\Omega}\e{m},\bar{\Omega}_{\Lambda})$ is the theoretical distance modulus of a source at redshift $z$, in a FL universe with cosmological parameters $\bar{\Omega}\e{m}$, $\bar{\Omega}_{\Lambda}$, $\bar{\Omega}_{K}=1-\bar{\Omega}\e{m}-\bar{\Omega}_{\Lambda}$.

\begin{figure}[!h]
\centering
\includegraphics[width=0.99\columnwidth]{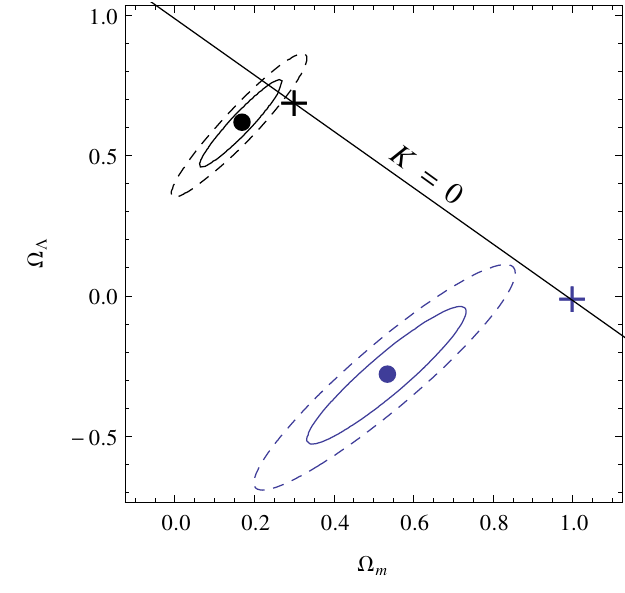}
\caption{Comparison between background parameters (crosses) and apparent parameters (disks) for two Swiss-cheese models with $f=f\e{min}$ and $M=10^{15} M_\odot$. In blue, $(\Omega\e{m},\Omega_{\Lambda})=(1,0)$ leads to $(\bar{\Omega}\e{m},\bar{\Omega}_{\Lambda})=(0.5,-0.3)$. In black, $(\Omega\e{m},\Omega_{\Lambda})=(0.3,0.7)$ leads to $(\bar{\Omega}\e{m},\bar{\Omega}_{\Lambda})=(0.2,0.6)$. The $1\sigma$ and $2\sigma$ contours are respectively the solid and dashed ellipses. The solid straight line indicates the configurations with zero spatial curvature.}
\label{fig:contours}
\end{figure}

The results of this analysis for two mock Hubble diagrams are shown in Fig. \ref{fig:contours}. An EdS background leads to apparent parameters $(\bar{\Omega}\e{m},\bar{\Omega}_{K},\bar{\Omega}_{\Lambda})=(0.5,0.8,-0.3)$, which are very different from $(1,0,0)$. Thus, the positive shift of $D\e{L}(z)$---clearly displayed in Fig.~\ref{fig:example_Mock_Hubble}---turns out to be mostly associated to an apparent spatial curvature, rather than to an apparent cosmological constant. In this case the apparent curvature is necessary to obtain a good fit ($\bar{\Omega}_K=0$ is out of the $2\sigma$ confidence contour), because a spatially flat FL model does not allow us to reproduce both the low-$z$ and high-$z$ behaviors of the diagram. The effect is weaker for a background with $(\Omega\e{m},\Omega_{\Lambda})=(0.3,0.7)$, which leads to $(\bar{\Omega}\e{m},\bar{\Omega}_K,\bar{\Omega}_{\Lambda})=(0.2,0.2,0.6)$.

\subsection{Quantitative results}
\label{subsec:results_cosmo_parameters}

\subsubsection{Influence of the smoothness parameter}
\label{subsubsec:varying_f}

\begin{figure}[!h]
\centering
\includegraphics[width=\columnwidth]{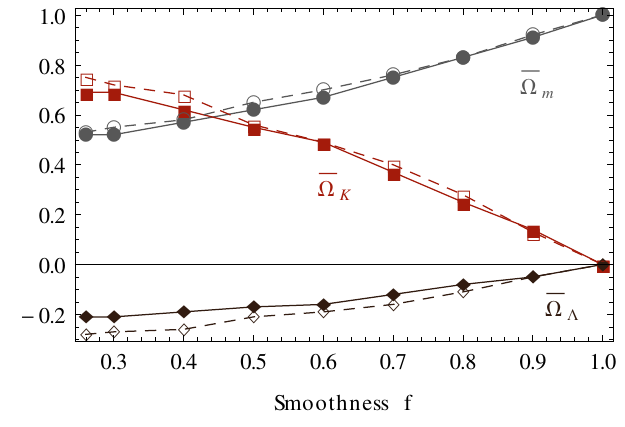}
\caption{Apparent cosmic parameters $\bar{\Omega}\e{m}$ (gray disks), $\bar{\Omega}_K$ (red squares) and $\bar{\Omega}\e{m}$ (black diamonds) versus smoothness parameter $f$, for a Swiss-cheese universe with EdS background $(\Omega\e{m},\Omega_K,\Omega_\Lambda)=(1,0,0)$. Solid lines and filled markers correspond to $M=10^{11} M_\odot$, dashed lines and empty markers to $M=10^{15} M_\odot$.}
\label{fig:cosmic_parameters_versus_f}
\end{figure}

Consider a Swiss-cheese model with EdS background cosmology. Figure~\ref{fig:cosmic_parameters_versus_f} shows the evolution of the apparent cosmological parameters with smoothness~$f$. As expected, we recover $\bar{\Omega}_i=\Omega_i$ when $f=1$, the discrepancy between background and apparent cosmological parameters being maximal when $f=f\e{min}$. Surprisingly, a Swiss-cheese universe seems progressively dominated by a negative spatial curvature for small values of $f$.

\begin{figure}[!h]
\centering
\includegraphics[width=\columnwidth]{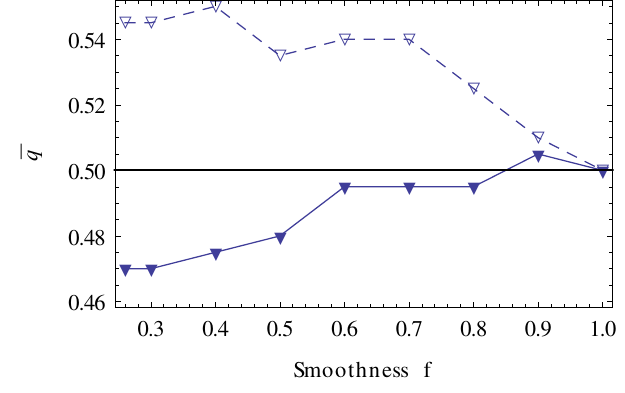}
\caption{Apparent deceleration parameter $\bar{q}$ as a function of smoothness parameter $f$, for Swiss-cheese models with EdS background ($q=1/2$). Solid lines and filled markers correspond to $M=10^{11} M_\odot$, dashed lines and empty markers to $M=10^{15} M_\odot$.}
\label{fig:q_versus_f}
\end{figure}

The apparent deceleration parameter $\bar{q}=\bar{\Omega}\e{m}/2-\bar{\Omega}_\Lambda$ is plotted in Fig.~\ref{fig:q_versus_f} as a function of $f$. Interestingly, even for $f=f\e{min}$, $\bar{q}$ remains almost equal to its background value $q=1/2$. Therefore, though the apparent cosmological parameters can strongly differ from the background ones, the apparent expansion history is almost the same---at second order---as the background one.

Note that the results displayed in Figs.~\ref{fig:cosmic_parameters_versus_f} and \ref{fig:q_versus_f} are consistent with each other. The apparent cosmological constant $\bar{\Omega}_\Lambda$ is slightly smaller for $M=10^{15}M_\odot$ than for $M=10^{11}M_\odot$, so that $\bar{q}$ is slightly larger.

\subsubsection{Influence of the background cosmological constant}
\label{subsubsec:varying_Lambda}

\begin{figure}[!h]
\centering
\includegraphics[width=\columnwidth]{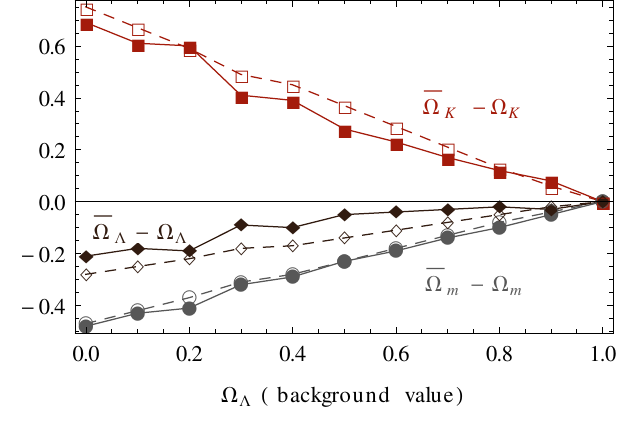}
\caption{Difference between apparent and background cosmological parameters $\bar{\Omega}\e{m}-\Omega\e{m}$ (gray disks), $\bar{\Omega}_K-\Omega_K$ (red squares) and $\bar{\Omega}_\Lambda-\Omega_\Lambda$ (black diamonds) versus background~$\Omega_\Lambda$, for Swiss-cheese models with $f=f\e{min}$. Solid lines and filled markers correspond to $M=10^{11} M_\odot$, dashed lines and empty markers to $M=10^{15} M_\odot$.}
\label{fig:cosmic_parameters_versus_Lambda}
\end{figure}

Now consider a Swiss-cheese model with $f=f\e{min}$ and change its background cosmology. Figure~\ref{fig:cosmic_parameters_versus_Lambda} shows the evolution of the apparent cosmological parameters versus the background cosmological constant $\Omega_\Lambda$. As it could have already been suspected from Fig. \ref{fig:contours}, the difference between apparent and background parameters decreases with $\Omega_{\Lambda}$, and vanishes in a de Sitter universe. This can be understood as follows. The construction of a Swiss-cheese universe consists in changing the spatial distribution of the pressureless matter, while the cosmological constant remains purely homogeneous. Thus, the geometry of spacetime is less affected by the presence of inhomogeneities if $\Omega_\Lambda/\Omega\e{m}$ is greater. In the extreme case $(\Omega\e{m},\Omega_\Lambda)=(0,1)$, any Swiss cheese is identical to its background, since there is no matter to be reorganized.

\begin{figure}[!h]
\centering
\includegraphics[width=\columnwidth]{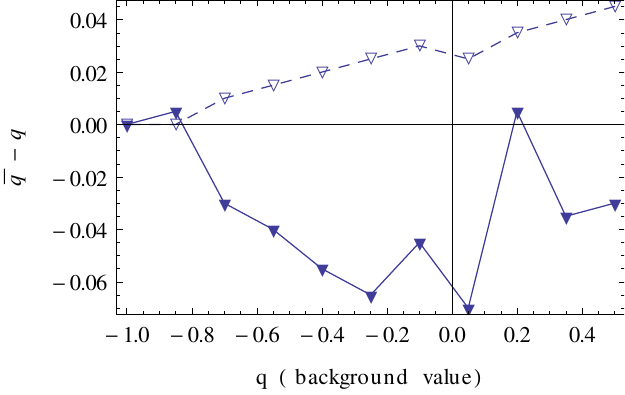}
\caption{Difference between apparent and background deceleration parameters~$\bar{q}-q$ as a function of $q$, for Swiss-cheese models with $f=f\e{min}$. Solid lines and filled markers correspond to $M=10^{11} M_\odot$, dashed lines and empty markers to $M=10^{15} M_\odot$.}
\label{fig:Deltaq_versus_q}
\end{figure}

We also plot in Fig.~\ref{fig:Deltaq_versus_q} the difference between the apparent deceleration parameter~$\bar{q}$ and the background one~$q=\Omega\e{m}/2-\Omega_\Lambda$, as a function of $q$. Again, $\bar{q}$ does not significantly differ from $q$. This result must be compared with Fig.~11 of Ref.~\cite{DR74}, where $(\bar{q}-q)/q \approx 100\%$.

\subsubsection{Comparison with other recent studies}

The impact of a modified luminosity-redshift relation---due to inhomogeneities---on the cosmological parameters has already been investigated by several authors. In Ref.~\cite{Marra:2007pm}, it has been suggested that a Swiss-cheese model with LTB holes and EdS background displays an apparent cosmological constant~$\bar{\Omega}_\Lambda=0.4$; but as already mentioned in Subsec.~\ref{subsec:summary_discussion}, such a claim was proven to be inaccurate in Refs.~\cite{Brouzakis:2008,Vanderveld:2008vi,Vanderveld:2011}, because it relies on observations along a peculiar line of sight. When many random directions of observation are taken into account, the mean magnification goes back to $1$. Hence, contrary to our results, the apparent cosmological parameters of a Swiss-cheese model with LTB holes are identical to the background ones. This conclusion is in agreement with Ref.~\cite{BolejkoFerreira2012}, where similar studies are performed in various cosmological toy models; and also with Ref.~\cite{Valageas2000} in the framework of perturbation theory.

However, it is crucial to distinguish those approaches (LTB Swiss-cheese models and perturbation theory) from the one adopted in this article, because they do not address the same issue. The former share the purpose of evaluating the influence of inhomogeneities smoothed on large scales, while we focused on smaller scales for which matter cannot be considered smoothly distributed. Thus, our results must not be considered different, but rather complementary.

\subsection{An alternative way to fit the Hubble diagram}
\label{subsec:alternative_fit}

Let us now address the converse problem, and determine the background cosmological parameters of the Swiss-cheese model that best reproduces the actual observations. For that purpose, the simplest method would be to fit our observed Hubble diagram using the theoretical luminosity-redshift relation $D\e{L}\h{SC}(z)$ of a Swiss-cheese universe. Hence, we need to derive such a relation in order to proceed.

\subsubsection{Analytical estimation of the distance-redshift relation of a Swiss-cheese universe}

As argued in Sec. \ref{sec:results_one_hole}, any observationally relevant light beam which crosses a Kottler region has an impact parameter $b$ much larger than the Schwarzschild radius $r\e{S}$ of the central object. Moreover, since the cosmological constant has no effect on light focusing, we conclude that inside a hole, the evolution of the cross-sectional area of a light beam behaves essentially as in Minkowski spacetime. This conclusion is supported by the perturbative calculation of the Wronski matrix $\boldsymbol{\mathcal{W}}\e{K}$ performed in \S~\ref{paragraph:perturbative_Wronski}.

If both the observer and the source are located on the surface of a hole, their angular distance is therefore $D\e{A}\h{hole} = \sqrt{\det \boldsymbol{\mathcal{D}}} \approx v\e{out}-v\e{in}$, where $v$ denotes the affine parameter. More generally, for a beam which crosses $N$ contiguous holes, we get
\begin{equation}
D\e{A}\h{holes} \approx \sum_{i=1}^{N} \Delta v_i,
\end{equation}
where $\Delta v_i \equiv v_{\text{out},i} - v_{\text{in},i}$ is the variation of the affine parameter between entrance into and exit from the $i$th hole. Let us now evaluate $\Delta v_i$. The time part of the geodesic equation in Kottler geometry yields
\begin{equation}
k^t \equiv \frac{\dd t}{\dd v} 
= \frac{E}{A(r)}
\overset{r \gg r\e{S}}{\approx} E = \text{constant},
\end{equation}
where $E$ is the usual constant of motion. We conclude that $\Delta v_i \approx k^t_{\text{out},i}\,\Delta t_i$. Besides, the relations \eqref{eq:relation_between_times} and~\eqref{eq:ktout} between FL and Kottler coordinates on the junction hypersurface, together with $A(r\e{h})\approx 1$, lead to $\Delta t_i \approx \Delta T_i$ and $k^t\e{out} \approx a\e{out}/a_0$. Finally,
\begin{equation}
D\e{A}\h{holes} \approx \sum_{i=1}^{N} \frac{a_{\text{out},i}}{a_0} \, \Delta T_i \approx \int_T^{T\e{obs}} \frac{a(T')}{a_0} \; \dd T' ,
\end{equation}
where we approximated the sum over $i$ by an integral. This operation is valid as far as $\Delta T_i$ remains small compared to the Hubble time. In terms of redshifts, we have
\begin{equation}
D\e{A}\h{holes}(z) = \int_0^z \frac{\dd z'}{(1+z')^2 H(z')}.
\end{equation}
By construction, this formula describes the behavior of the angular distance when light only travels through Kottler regions. In order to take the FL regions into account, we write the distance-redshift relation~$D\e{A}\h{SC}(z)$ of the Swiss cheese as the following (heuristic) linear combination
\begin{equation}
D\e{A}\h{SC}(z) = (1-f)\,D\e{A}\h{holes}(z) + f\,D\e{A}\h{FL}(z),
\label{eq:analytical_DA_z_SC}
\end{equation}
where $f$ still denotes the smoothness parameter defined in \S~\ref{subsubsec:smoothnes_parameter}, and $D\e{A}\h{FL}(z)$ is the distance-redshift relation in a FL universe, given by Eq. \eqref{eq:FL_DA-z_relation}.

\begin{figure}[!h]
\centering
\includegraphics[width=\columnwidth]{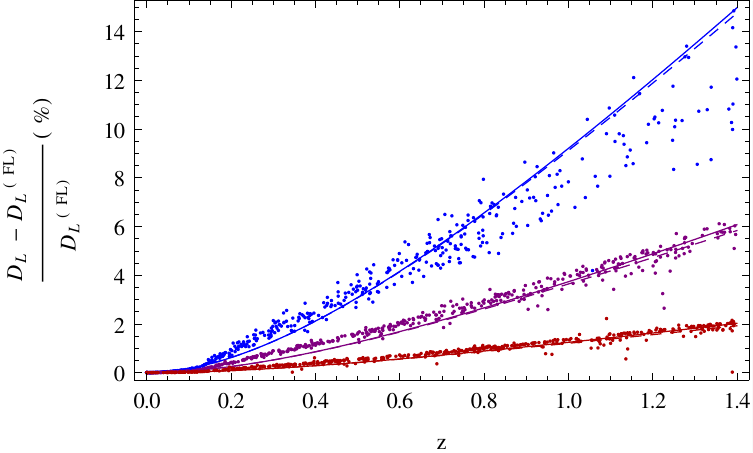}
\caption{Comparison between the approximate luminosity-redshift relation $D\e{L}\h{SC}(z)=(1+z)^2 D\e{A}\h{SC}(z)$ in a Swiss-cheese universe (solid lines), simulated observations (dots), and the Dyer-Roeder model~$D\e{L}\h{DR}(z)$ with $\alpha(z)=f$ (dashed lines). Three different values of the smoothness parameter are tested, from top to bottom: $f=f\e{min}\approx0.26$, $f=0.7$, $f=0.9$.}
\label{fig:analytical_estimation}
\end{figure}

A comparison between the above analytical estimation and the numerical results is plotted in Fig. \ref{fig:analytical_estimation}. The agreement is qualitatively good, especially as it is obtained without any fitting procedure. Moreover, it is straightforward to show that $D\e{A}\h{SC}(z)$ and $D\e{A}\h{FL}(z)$ are identical up to second order in $z$. This is in agreement with---and somehow explains---the numerical results of \S~\ref{subsubsec:varying_f} and \S~\ref{subsubsec:varying_Lambda}, where we showed that the apparent deceleration parameter~$\bar{q}$ is the same as the background one $q$.

Note finally that the general tendency of our analytical relation is to overestimate $\delta D\e{L}$ at high redshifts. The main reason is that its derivation uses the behavior of $\boldsymbol{\mathcal{W}}\e{K}$ at zeroth order in $r\e{S}/b$; that is, it neglects the effect of the central mass in the Kottler region. The first-order term in $\boldsymbol{\mathcal{W}}\e{K}$---taken into account in the numerical results---tends to lower the associated luminosity distance.

\subsubsection{Comparison with the Dyer-Roeder approach}

Another widely used approximation to model the propagation of light in underdense regions was proposed by Dyer and Roeder \cite{DR72} in 1972. It assumes that (1)~the Sachs equation and the relation $v(z)$ are the same as in a FL spacetime---in particular, the null shear vanishes---and (2)~the optical parameter $\Phi_{00}$ (see \S~\ref{subsubsec:Sachs_equation}) is replaced by $\alpha(z)\Phi_{00}$, where $\alpha(z)$ represents the fraction of matter intercepted by the geodesic bundle. In brief, the DR model encodes that light propagates mostly in underdense regions by reducing the Ricci focusing, while still neglecting the Weyl focusing. Under such conditions, the DR expression of the angular distance $D\e{A}\h{DR}(z)$ is determined~by
\begin{multline}
\frac{\dd^2 D\e{A}\h{DR}}{\dd z^2} + \left(\frac{\dd\ln H}{\dd z} +\frac{2}{1+z}\right) \frac{\dd D\e{A}\h{DR}}{\dd z} \\
 =-\frac{3\,\Omega\e{m}}{2}
 \left( \frac{H_0}{H} \right)^2 
 (1+z)\,\alpha(z)\,D\e{A}\h{DR}(z).
\label{eq:dr-bgd0}
\end{multline}
This attempt to model the average effect of inhomogeneities, while assuming that the dynamics of the Universe is the same as an isotropic and homogeneous spacetime, has been widely questioned~\cite{ehlerss,sasaki,Tomita:1999tg,Rasanen:2008be} and recently argued to be mathematically inconsistent~\cite{cemuu}. 

Interestingly, our estimation $D\e{A}\h{SC}(z)$ of the distance-redshift relation in a Swiss-cheese universe reads
\begin{multline}
\frac{\dd^2 D\e{A}\h{SC}}{\dd z^2} + \left(\frac{\dd\ln H}{\dd z} +\frac{2}{1+z}\right) \frac{\dd D\e{A}\h{SC}}{\dd z} \\
 =-\frac{3\,\Omega\e{m}}{2}
 \left( \frac{H_0}{H} \right)^2 
 (1+z)\,f\,D\e{A}\h{FL}(z).
\end{multline}
which is similar to Eq.~\eqref{eq:dr-bgd0} with $\alpha(z)=f$, except that the right-hand side is proportional to $D\e{A}\h{FL}$ instead of~$D\e{A}\h{SC}$. Nevertheless, it turns out that such a difference has only a very weak impact, in the sense that
\begin{equation}
D\e{A}\h{SC}(z) \approx D\e{A}\h{DR}(z),
\quad \text{i.e.}  \quad
D\e{L}\h{SC}(z) \approx D\e{L}\h{DR}(z),
\end{equation}
if $\alpha(z)=f$. This appears clearly in Fig. \ref{fig:analytical_estimation}, where the dashed and solid lines are almost superimposed. In fact, it is not really surprising, since both approaches rely on the same assumptions: no backreaction, no Weyl focusing and an effective reduction of the Ricci focusing.

Note however that this approach models the effect of the inhomogeneities on the mean value of the luminosity distance but does not address the dispersion of the data.

\subsubsection{Fitting real data with $D\e{L}\h{SC}(z)$}

The modified luminosity-redshift relation~$D\e{L}\h{SC}(z)$ derived in the previous paragraph can be used to fit the observed Hubble diagram. We apply the same $\chi^2$ method as described in Subsec.~\ref{subsec:determining_cosmological_parameters}, except that now (1) the triplets $(z_i,\mu_i,\Delta\mu_i)$ are observations of the SNLS 3 catalog~\cite{SNLS3}, and (2) $\mu\e{FL}(z \,|\, \bar{\Omega}\e{m},\bar{\Omega}_{\Lambda})$ is replaced by $\mu\e{SC}(z \,|\, \Omega\e{m},f)$, where the background curvature~$\Omega_K$ is fixed to $0$ (so that $\Omega_\Lambda=1-\Omega\e{m}$). Hence, we are looking for the smoothness parameter~$f$, and the background cosmological parameters, of the spatially Euclidean Swiss-cheese model which best fits the actual SN observations.

\begin{figure}[!h]
\includegraphics[width=\columnwidth]{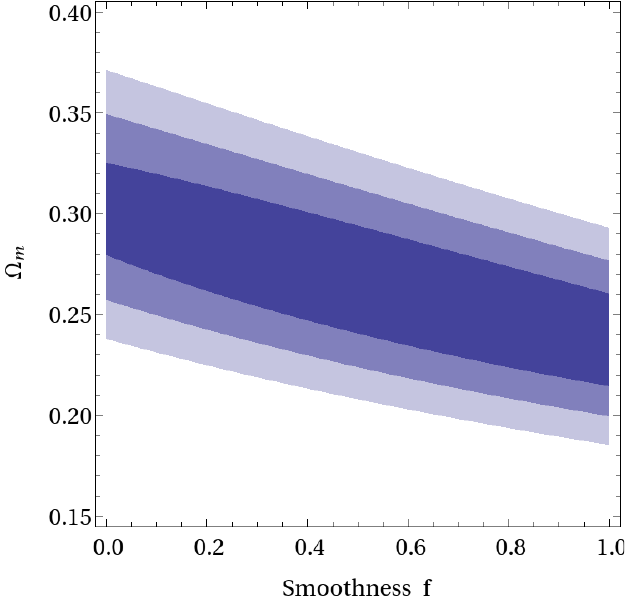}
\caption{Fit of the Hubble diagram constructed from the SNLS 3 data set \cite{SNLS3}, by using the luminosity-redshift relation $D\e{L}\h{SC}(z\,|\,\Omega\e{m},f)$ of a spatially Euclidean Swiss-cheese model. The colored areas indicate (from the darkest to the lightest) the $1\sigma$, $2\sigma$ and $3\sigma$ confidence levels.}
\label{fig:alternative_fit}
\end{figure}

The results of the $\chi^2$ fit are displayed in Fig.~\ref{fig:alternative_fit}. First of all, we note that the confidence areas are very stretched horizontally, so that the smoothness parameter~$f$ cannot be reasonably constrained by the Hubble diagram. There are two reasons for this. On the one hand, we know from \S~\ref{subsubsec:varying_Lambda} that $f$ has only a weak influence on the luminosity-redshift relation in a universe dominated by the cosmological constant, which is the case here ($\Omega_\Lambda \sim 0.7$--$0.8$). On the other hand, since $D\e{L}\h{SC}(z)$ and $D\e{L}\h{FL}(z)$ only differ by terms of order $z^3$ and higher, one would need more high-redshift observations to discriminate them. However, all the current SNe catalogs---including the SNLS 3 data set---contain mostly low-redshift SNe.

Besides, Fig.~\ref{fig:alternative_fit} shows that fixing a given value of $f$ changes the best-fit value of $\Omega\e{m}$. In the extreme case of a Swiss cheese only made of clumps ($f=0$) we get $\Omega\e{m}=0.3$, while in the FL case $(f=1)$ the best value is $\Omega\e{m}=0.24$, in agreement with Ref.~\cite{SNLS3}. Such a discrepancy, of order $20\,\%$, is significant in the era of precision cosmology, where one aims at determining the cosmological parameters at the percent level.

Let us finally emphasize that such a fit is only indicative, because it relies on an approximation of the luminosity-redshift relation in a Swiss-cheese universe.

\section{Conclusion}
\label{sec:conclusion}

In this article, we have investigated the effect of the distribution of matter on the Hubble diagram, and on the resulting inference of the cosmological parameters. For that purpose, we have studied light propagation in Swiss-cheese models. This class of exact solutions of the Einstein field equations is indeed very suitable, because it can describe a strongly inhomogeneous distribution of matter which does not backreact on the global cosmic expansion. The latter is entirely governed by the background cosmological parameters~$\Omega_{\rm m}$,~$\Omega_{\Lambda}$ characterizing the FL regions of the model. The inhomogeneities are clumps of mass~$M$, while the fraction of remaining fluid matter is $f$---called smoothness parameter. The Swiss-cheese models are therefore defined by two ``dynamical'' parameters $(\Omega_{\rm m},\Omega_{\Lambda})$, and two ``structural'' parameters~$(f,M)$.

The laws of light propagation in a Swiss-cheese universe have been determined by solving the geodesic equation and the Sachs equation. For the latter, we have introduced a new technique---based on the Wronski matrix---in order to deal more easily with a patchwork of spacetimes. Our results, mostly analytical, have been included in a Mathematica program, and used to compute the impact of the Swiss-cheese holes on the redshift and on the luminosity distance. For a light beam which crosses many holes, we have shown that the effect on the redshift remains negligible, while the luminosity distance increases significantly with respect to the one observed in a FL universe ($\delta D\e{L} \sim 10\,\%$ for sources at $z \sim 1$), inducing a bias in the Hubble diagram.

The consequences of the bias on the inference of the cosmological parameters have been investigated by simulating Hubble diagrams for various Swiss-cheese models, and by fitting them with the usual FL luminosity-redshift relation. In general, the resulting ``apparent'' cosmological parameters are very different from the ``background'' ones which govern the cosmic expansion, but in a way that leaves the deceleration parameter unchanged. Moreover, the discrepancy between apparent and background cosmological parameters turns out to decrease with $\Lambda$, and is therefore small for a universe dominated by the cosmological constant. Finally, we have derived an approximate luminosity-redshift relation for Swiss-cheese models, which is similar to the one obtained following the Dyer-Roeder approach. Using this relation to fit the Hubble diagram constructed from the SNLS 3~data set, we have found that the smoothness parameter cannot be constrained by such observations. However, turning arbitrarily $f=1$ into $f=0$ has an impact of order $20\,\%$ on the best-fit value of $\Omega\e{m}$, which is significant in the era of precision cosmology (see Ref.~\cite{SC_Planck} for further discussion).

Of course, our model is oversimplifying for various reasons. First, it does not take into account either the complex distribution of the large scale structures, or the presence of diffuse matter on small scales---such as gas and possibly dark matter. Second, it does not take strong lensing effects into account, assuming that clumps are ``opaque''. We can conjecture that this overestimates the actual effect of the inhomogeneities. Nevertheless, it shows that their imprint on the Hubble diagram cannot be neglected, and should be modeled beyond the perturbation regime. Note finally that several extensions are allowed by our formalism. For instance, we could introduce different kinds of inhomogeneities, in order to construct fractal structures for which the smoothness parameter is arbitrarily close to zero. Additionally to the Hubble diagrams, we could also generate the shear maps of Swiss-cheese models, and determine whether their combination allows for better constraints on the various parameters.

This work explicitly raises the question of the meaning of the cosmological parameters, and of whether the values we measure under the hypothesis of a pure FL background represent their ``true'' or some ``dressed'' values. Similar ideas have actually been held in other contexts~\cite{carfora,wiltshire}, and in particular regarding the spatial curvature \cite{Clarkson2011,Coley2005}. We claim that the simplest Swiss-cheese models are good models to address such questions---as well as the Ricci-Weyl problem and the fluid approximation---with their own use, between the perturbation theory and $N$-body simulations.

\section*{Acknowledgements}

We thank  Francis Bernardeau, Thomas Buchert, Nathalie Palanque-Delabrouille, Peter Dunsby, George Ellis, Yannick Mellier, Shinji Mukohyama, Cyril Pitrou, Carlo Schimd, and Markus Werner for discussions. We also thank Julian Adamek, Krzysztof Bolejko, Alan Coley, Pedro Ferreira, Giovanni Marozzi and Zdenek Stuchl\'ik for comments. PF thanks the \'Ecole Normale Sup\'erieure de Lyon for funding during his master and first year of Ph.D. HD thanks the Institut Lagrange de Paris for funding during her master's and the Institut d'Astrophysique de Paris for hospitality during the later phase of this work. This work was supported by French state funds managed by the ANR within the Investissements d'Avenir programme under reference ANR--11--IDEX--0004--02 and the Programme National Cosmologie et Galaxies.


\end{document}